\documentclass[10pt,draftcls,onecolumn]{IEEEtran}
\usepackage{cite}
\usepackage{amsmath, amssymb}
\usepackage{latexsym,wasysym,marvosym}
\DeclareMathAlphabet{\mathbbm}{U}{bbm}{m}{n}% from bbm.sty
\usepackage{amssymb,stmaryrd}
\usepackage{graphicx,color}
\usepackage{psfig}
\usepackage{subfigure}
\usepackage{url}
\usepackage{array,multirow}
\usepackage{fontenc,times}
\usepackage{pxfonts}
\usepackage{dsfont}
\makeatletter

\newtheorem{theorem}{\textbf{Theorem}}

\newtheorem{proposition}[theorem]{\textbf{Proposition}}
\newtheorem{example}{\textbf{Example}}
\newtheorem{definition}{\textbf{Definition}}
\newtheorem{lemma}[theorem]{\textbf{Lemma}}
\newtheorem{remark}{\textbf{Remark}}

\newcommand{\mc}{\mathcal} %

\newcommand{\mkv}{- \!\!\! \minuso \!\!\! -}

\IEEEoverridecommandlockouts

\usepackage{psfrag, xcolor}
\usepackage{hyperref}
\usepackage{enumitem}
\usepackage{cite,epsfig,url}
\usepackage{booktabs}
\allowdisplaybreaks[4]

\makeatletter
\def\IEEElabelanchoreqn#1{\bgroup
	\def\@currentlabel{\p@equation\theequation}\relax
	\def\@currentHref{\@IEEEtheHrefequation}\label{#1}\relax
	\Hy@raisedlink{\hyper@anchorstart{\@currentHref}}\relax
	\Hy@raisedlink{\hyper@anchorend}\egroup}
\makeatother
\newcommand{\subnumberinglabel}[1]{\IEEEyesnumber
	\IEEEyessubnumber*\IEEElabelanchoreqn{#1}}

\newlength{\eqboxstorage}

\newcommand{\W}{\mathsf{W}}

\begin{document}
	\fontencoding{OT1}\fontsize{10}{11}\selectfont
	
	\title{Distributed Hypothesis Testing: Cooperation and Concurrent Detection}

	\author{Pierre Escamilla$^{\dagger}$$^{\star}$  \qquad Mich\`ele Wigger $^{\star}$ \qquad Abdellatif Zaidi$^{\dagger}$ $^{\ddagger}$\vspace{0.3cm}\\
		$^{\dagger}$ Paris Research Center, Huawei Technologies, Boulogne-Billancourt, 92100, France\\
		$^{\ddagger}$ Universit\'e Paris-Est, Champs-sur-Marne, 77454, France\\
		$^{\star}$ LTCI, T\'el\'ecom ParisTech, Universit\'e Paris-Saclay, 75013 Paris, France\\
		\{\tt pierre.escamilla@huawei.com, michele.wigger@telecom-paristech.fr\}\\
		\{\tt abdellatif.zaidi@u-pem.fr\}
		%\vspace{-5mm}
	}
	
	\maketitle
	\begin{abstract}
%		One  detection system with a single Sensor and two Detectors is considered, where each of the terminals observes a memoryless source sequence, the  Sensor sends a message  to both Detectors, and the first Detector also send a message to the second Detector. Communication of these messages is assumed to be error-free but rate-limited. The joint probability mass function (pmf) of the  source sequences observed at the three terminals depends on an $\M$-ary hypothesis $(\M \geq 2)$, and the goal of the communication is that each Detector can guess the underlying hypothesis. Detector $1$ and Detector $2$ aim to maximize the error exponent \textit{under hypothesis} $i_k$, $i_k \in \{1,\ldots,\M\}$, while ensuring a small probability of error under all other hypotheses. We study this problem in the case in which the Detectors aim at maximizing their error exponents under the \textit{same} hypothesis (i.e., $i_1=i_2$) and in the case in which they aim at maximizing their error exponents under \textit{distinct} hypotheses (i.e., $i_1 \neq i_2$). For this setting and the exact same setting but where communication from first Detector is not allowed we present achievable error exponents region for the case of positive communication rates. We also specialize the results to some specific cases of testing against independence and zero rate communication for which we characterize the optimal exponents region.
		A single-sensor two-detectors system is considered where the sensor communicates with both detectors and Detector 1 communicates with Detector 2, all over noise-free rate-limited links. The sensor and both detectors observe discrete memoryless source sequences whose  joint probability mass function  depends on a binary hypothesis. The goal at each detector is to guess the binary hypothesis in a way that, for increasing observation lengths, the  probability of error under one of the hypotheses %(the so called Type-II error probability) 
		decays to zero with largest possible exponential decay, whereas the  probability of error under the other hypothesis %(the so called Type-I error probability)
		 can decay to zero arbitrarily slow. For the setting with zero-rate communication on both links, we exactly characterize the set of  possible exponents and the gain brought up by  cooperation, in function of the  number of bits that are sent over the two links. Notice that, for this setting, tradeoffs between the exponents achieved at the two detectors arise only in few particular cases. In all other cases, each detector achieves the same performance as if it were the only detector in the system. %Comparing the new to our previous results allows us to exactly quantify the benefits of cooperation in this zero-rate scenario. 
		For the setting with positive communication rates from the sensor to the detectors, we characterize the set of all possible exponents in  a special case of testing against independence. In this case the cooperation link allows Detector~2 to increase its Type-II error exponent by an amount that is equal to the exponent attained at Detector 1. We also provide a general inner bound on the set of achievable error exponents. For most cases it shows a tradeoff between the two exponents. % the  the detectors aim to guess.   %Detector 1 and Detector 2 aim to maximize the error exponent under hypothesis $i_k$ , $i_k \in {1,2}$, while ensuring a small probability of error under all other hypotheses. We study this problem in the case in which the Detectors aim at maximizing their error exponents under the same hypothesis (i.e., $i_1$ = $i_2$ ) and in the case in which they aim at maximizing their error exponents under distinct hypotheses (i.e., $i_1 \neq i_2$ ). For this setting and the exact same setting but where communication from first Detector to second Detector is not allowed we present achievable error exponents region for the case of positive communication rates. We also specialize the results to some specific cases of testing against independence and zero rate communication for which we characterize the optimal exponents region.
	\end{abstract}
	
%------------------------------------------------------------------------------------------------------------------------------------
	\section{Introduction}~\label{secI}
			
	\begin{figure}[!t]
		\begin{center}
			\includegraphics[width=0.8\linewidth]{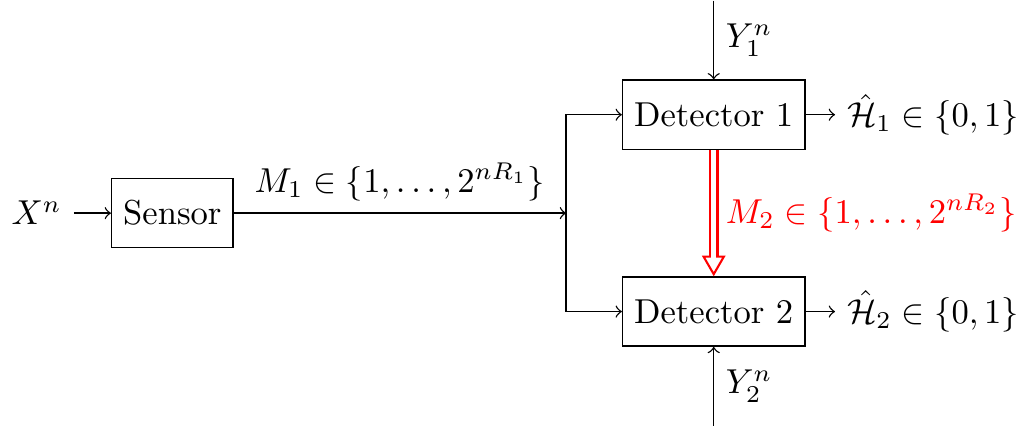}
			\caption{A Heegard-Berger type source coding model with unidirectional conferencing for multiterminal hypothesis testing.}
			\label{fig-system-model}
		\end{center}
		\vspace{-5mm}
	\end{figure}
	
	%------------------------------------------------------------------------------------------------------------------------------------

	Problems of distributed hypothesis testing are strongly rooted in both statistics and information theory. In particular,  \cite{AC86, H87,SHA94} considered a distributed hypothesis testing problem where a  a single sensor communicates with  a single detector over a rate-limited but noise-free link. The goal of \cite{AC86,H87} and \cite{SHA94} was to determine the largest Type-II error exponent under a fixed constraint on the Type-I error exponent.  Ahlswede and Csisz\'ar  in \cite{AC86} presented a coding and testing scheme and the corresponding Type-II error exponent for this problem, and established optimality of the exponent for the special case of \emph{testing against independence}. For the general case,  the scheme was subsequently improved by   Han  \cite{H87} and by Shimokawa, Han, Amari  \cite{SHA94}. The latter scheme was shown to achieve the optimal exponent  in the special case of \emph{testing against conditional independence} by Rahman and Wagner \cite{RW12}.  
	
	This line of works has also been extended to networks with multiple sensors \cite{H87,RW12,ZL14A,ZE19,UEZ18}, multiple detectors \cite{SWT18}, interactive terminals~\cite{TC08,XK12,KPD16A}, multi-hop networks~\cite{ZL14B,ZL15,WT16,SWW17,EWZ18},  noisy channels~\cite{SW18A,SG17} and scenarios with privacy constraints~\cite{LSCV17,LSTC18,SGC18,GBST19}. The works most closely related to the present manuscript are \cite{WT16,SWT18} and \cite{ZL18}. The former two, \cite{WT16,SWT18}, fully characterize  the set of possible Type-II error exponents in the special case of testing against independence and testing against conditional independence for a scenario  with a single sensor and two non-cooperative detectors. 
 In this paper we consider a similar scenario but where one of the detectors can send a cooperation message to the other detector. The same configuration was already considered in  \cite{ZL18} for a specific case of testing against independence (see Remark~\ref{rem:zhou-lai} in Section~\ref{secII}) and under the assumption that only the detector receiving the cooperation message has to take a decision.

	The results discussed so far concern scenarios where  communication is of positive rates.
An important line of work also assumes zero rate communication.  The single-sensor single-detector version of this problem was addressed in 	
	   \cite{H87}  and~\cite{SP92}, where Han identified the optimal exponent for the case where only  a single bit is communicated and Shalaby and Papamarcou proved that this  exponent is also optimal when communication comprises a sublinear number of bits. The finite length regime was investigated in \cite{W18}. Zero-rate hypothesis testing in interactive setup and multi-hop networks were addressed respectively in \cite{KPD16B} and \cite{ZL15,EWZ18,EZW19}. In particular, in our previous work \cite{EWZ18}, we proved a similar result for the single-sensor two-detectors setup. In the current manuscript we consider the extension to a cooperative setup.  %work to the single- {\color{red}Are there other network extensions of zero-rate communication?} % Key to the derivation of the converse proof in~\cite{H87} is an ingenious use of the ``Blowing-Up" lemma~\cite[Theorem 5.4]{C11}. This lemma plays a similar crucial role for establishing converse parts for more general zero-rate hypothesis testing systems with exponential-type constraint on the Type I error \cite{HK89}. 

	%------------------------------------------------------------------------------------------------------------------------------------
%	\subsection{Focus and Main Contributions}

Specifically, in this paper we consider the single-sensor two-detectors system in Fig.~\ref{fig-system-model} where Detector~1, after receiving a message from the Sensor, can send a message to Detector 2. This additional message allows the detectors to collaborate in their decision and one of the the goal of the paper is to quantify the increase in the Type-II error exponents  that is enable by this collaboration. We show that even a single bit of communication between the detectors (the temptative guess about the hypothesis at the transmitting detector)  can provide an unbounded gain in the Type-II error exponent of the detector receiving the bit. 

On a more technical level, 
 the presence of a cooperation link between the detectors seems to make the problem of identifying the optimal Type-II exponents significantly more difficult. For example, without cooperation, the set of achievable exponents for testing against independence  has been  solved in \cite{WT16}, and it is achievable with a simple scheme that does not rely on binning. With  cooperation, we managed to identify the optimal exponents  only  under the additional assumption that the observations at the two detectors are independent under both hypotheses and the cooperation rate is zero. In the general case, binning is necessary, which makes it hard to prove optimality of the achieved exponent. Notable exceptions are the results included in \cite{RW12,SWT17,SW18B,SG17,ZE19}.% a proof of optimality of the atta to make the identification of the optimal error exponent    hard. 
 %In fact, when applying binning in a multi-sensors  or multi-detectors setup, the expressions characterizing the achieved exponents become very complicated, with many competing exponents, see e.g., \cite{SWW17}.  

   For the sake of simplicity, in the case of positive communication rates we therefore only present and analyze a simple coding scheme without binning and also without Heegard-Berger \cite{HB85} coding. We  prove that this simple scheme is optimal in a special case of testing against independence where it achieves an exponent at Detector 2  equal to the sum of the exponents at both detectors in a non-cooperative setup. Collaboration between detectors thus allows to accumulate the error exponents at the  detectors. The testing against independence problem considered in this paper differs from the one in \cite{ZL18}, where the first detector cannot achieve a positive error exponent.

In our two-detectors setup where each detector aims at maximizing the error exponent under one of the two hypotheses, two cases  can be distinguished: both detectors aim at maximizing their exponents under the same hypothesis (we refer to this setup as \emph{coherent detection}) or the two detectors aim at maximizing their exponents under different hypotheses (we refer to this setup as \emph{concurrent detection}). In this paper we consider both scenarios. The exponents region can significantly differ under the two, in particular when based on its own observation the sensor  can guess the hypothesis, communicate this guess to the detectors, and adapt the communication to this guess. With  this strategy, the exponents region achieved by our simple scheme is a rectangle under concurrent detection, which means that each detector's exponent is the same as in a setup where the other detector is not present. Under coherent detection or concurrent detection when the sensor cannot distinguish the two hypotheses,  the exponents region achieved by our scheme shows a tradeoff between the two exponents. These results are for positive communication rates.

%The described results assume a communication link of positive rate from the sensor to the detectors. 

We also consider the case with fixed-length communication. Under coherent detection or concurrent detection when the sensor can send more than a single bit or cannot distinguish the two hypotheses,  the exponents region is a rectangle. In these cases, each detector  achieves the same exponent as if it were the only detector in the system. In contrast, a tradeoff arises if the sensor can distinguish the two hypotheses but can only send a single bit to the detectors. A comparison with the optimal exponents regions without cooperation  \cite{EWZ18},  allows us to exactly quantify the benefits of  detector collaboration in this setup with fixed communication alphabets. All results explained in this paragraph remain valid when the alphabets size are not fixed but grow sublinearly in the length of the observed sequences. They also generalize to an arbitrary number of hypotheses. Whereas for two detectors a tradeoff between the exponents arises only when the sensor sends a single bit to the detectors, in a multi-hypothesis testing scenario with $\mathsf{H} \geq 3$ such a tradeoff can arise whenever the number of communicated bits does not exceed $\log_2 \mathsf{H}$.

	\subsection{Paper Organization}
	
	The remainder of this paper is organized as follows. Section~\ref{secII} describes the system model. Sections~\ref{secIII} and Section~\ref{sec:moderate} describe our main results: Section~\ref{secIII} focuses on fixed communication alphabets and Section~\ref{sec:moderate} when communication is of positive rates. For the purpose of comparison, Section~\ref{sec:large} presents results for the  extreme case when  communication rates are very high. Technical proofs are referred to appendices. The paper is  concluded in Section~\ref{sec:conclusion}. 
\subsection{Notation}	
	Throughout, we use the following notation. Random variables are denoted by capital letters and their realizations by lower case, e.g., $X$ and $x$. A random or deterministic indexed $n$-tuple $X_1,\ldots, X_n$ or $x_1,\ldots, x_n$ is abbreviated as $X^n$ or as $x^n$. The set of all possible types of $n$-length sequences over $\mathcal{X}$ is denoted  $\mathcal{P}^n(\mathcal{X})$.
	For $\mu>0$, the set of sequences $x^n$ that are $\mu$-typical with respect to the pmf $P_X$ is denoted  $\mc T^n_{\mu}(P_X)$, the type of a sequence $x^n$ is denoted  by $\text{tp}(x^n)$. % and $\hat H (x^n)$.
	 The set of all $n$-length sequences with a given type $P_X$, or type class, is denoted $\mc T^n_{0}(P_X)$.  For random variables $X$ and $\bar{X}$ over the same alphabet $\mathcal{X}$  with pmfs $P_X$ and $\bar P_{X}$  satisfying $P_X \ll\bar P_{X}$ (i.e., for every $x \in \mathcal{X}$, if $P_{X} > 0 $ then also $\bar P_{X} > 0$), both $D(P_X\|\bar P_{X})$ and $D(X\|\bar{X})$ denote the Kullback-Leiber divergence between $X$ and $\bar{X}$. Finally, $H(\cdot)$ and $I(\cdot;\cdot)$ denote entropy and mutual information. %Where $H_P(A)$ and $I_P(A;B)$ indicate respectively that  $H(A)$ and $I(A;B)$ are computed according to $P$. %for a sequence $x_n$ the limit superior is 
%	\begin{equation}
%	\varlimsup_{n\to \infty}{x_n} = \lim_{n \to \infty} \big ( \sup_{m>n}{x_m} \big ),
%	\end{equation} and the limit inferior is
%	\begin{equation}
%	\varliminf_{n\to \infty}{x_n} = \lim_{n \to \infty} \big ( \inf_{m>n}{x_m} \big )
%	\end{equation}
	
%------------------------------------------------------------------------------------------------------------------------------------
\section{\textbf{Formal Problem Statement}}~\label{secII} %(from Allerton ’18)
	
Consider a three-terminal  problem with a Sensor observing the sequence $X^n$, a Detector 1 observing $Y_1^n$, and a Detector 2 observing $Y_2^n$. The joint probability mass function (pmf) of the tuple $(X^n,Y^n_1,Y^n_2)$ depends on one of two hypotheses. Under hypothesis 
\begin{equation}
\mathcal{H}=0 \colon \qquad \{(X_t,Y_{1,t}, Y_{2,t})\}_{t=1}^n \textnormal{ i.i.d. } P_{XY_1Y_2}
\end{equation} 
and under 
hypothesis 
\begin{equation}
\mathcal{H}=1 \colon \qquad \{(X_t,Y_{1,t}, Y_{2,t})\}_{t=1}^n \textnormal{ i.i.d. } \bar{P}_{XY_1Y_2}
\end{equation} 
   
The Sensor applies an encoding function
	\begin{equation}\label{eq:phi1}
	\phi_{1,n}\colon  \mc X^n \rightarrow \mc M_1\triangleq\{0,1,\ldots, \mathsf{W}_{1,n}-1\}
	\end{equation} 
	to its observed source sequence $X^n$ and sends the  resulting index 
	\begin{equation}
	M_1 = \phi_{1,n}(X^n)
	\end{equation}
	to both decoders. 
	Detector 1 then applies two functions to the pair $(M_1, Y_1^n)$ an encoding function 
	\begin{equation}\label{eq:phi2}
	\phi_{2,n} \colon \mc M_1 \times \mc Y^n_1 \rightarrow \mc M_2 \triangleq\{0,1,\ldots,\mathsf{W}_{2,n}-1\},
	\end{equation}
	and a decision function 
	\begin{equation}\label{eq:psi1}
	\psi_{1,n} \colon \mc M_1 \times \mc Y^n_1 \rightarrow \{0,1\}. % \{\mc H, \bar{\mc H}\}. 
	\end{equation}
	It sends the index 
	\begin{equation}\label{eq:M2}
	M_2=\phi_{2,n} (M_1, Y_1^n)
	\end{equation} to Detector 2, and decides on the hypothesis 
	\begin{equation}\hat{\mc H}_1 \triangleq \psi_{1,n} (M_1, Y_1^n).
	\end{equation}

	 Detector  2 applies a decision function 
	\begin{equation}\label{eq:psi2}
	\psi_{2,n}\colon  \mc M_1 \times \mc M_2 \times \mc Y^n_2 \rightarrow \{0,1\} % \{\mc H, \bar{\mc H}\}
	\end{equation}
	to the triple $(M_1, M_2, Y_2^n)$ to produce the decision  
	\begin{equation}\hat{\mc H}_2 \triangleq  \psi_{2,n} (M_1, M_2, Y_2^n).
	\end{equation} 
	
Both Detectors are required to have vanishing probabilities of error under both hypotheses. Moreover, for Detector 2, we require that the probability of error under $\mathcal{H}=1$ decays exponentially fast with the largest possible exponent.  For Detector 1, we consider two scenarios:  \textit{coherent detection} and \textit{concurrent detection}.  Under coherent detection, Detector 1 wishes to maximize the exponential decay of the probability of error under $H=1$. Under concurrent detection,  Detector 1 wishes to maximize the exponential decay of the probability of error under $H=0$. In an unifying manner, we define,  for $h_1 \in \{0,1\}$ and   $\bar{h}_1 = (h_1 + 1) \textnormal{ mod }2$, the following error probabilities:
	\begin{IEEEeqnarray}{rCl}
	\alpha_{1,n} &:= &\text{Pr}\big\{\hat{\mc H}_1 =\bar{h}_1   \big| \mc H = h_1\big\},\: \\ 
	\beta_{1,n}  & :=& \text{Pr}\big\{\hat{\mc H}_1 = h_1 \big| \mc H=\bar{h}_1 \big\},\: \\
	\alpha_{2,n} & :=&   \text{Pr}\big\{\hat{\mc H}_2 = 1 \big|\mc H = 0 \big\},\: \\ 
	 \beta_{2,n}  & :=& \text{Pr}\big\{\hat{\mc H}_2 = 0 \big|\mc H = 1  \big\}. 
	\end{IEEEeqnarray}
		
	\begin{definition}[Achievability under Rate-Constraints]%[Achievable Type-II Error Exponents under Constant Constraints on Type I Errors] 
		Given $\bar{h}_1\in\{0,1\}$ and rates $R_1, R_2 \geq 0$, % and small positive numbers $\epsilon_1, \epsilon_2 \in (0,1)$
		 an error-exponents pair $(\theta_1,\theta_2)$ is said achievable if for  all  blocklengths $n$ there exist  functions $\phi_{1,n}$, $\phi_{2,n}$, $\psi_{1,n}$ and $\psi_{2,n}$ as in \eqref{eq:phi1}, \eqref{eq:phi2}, \eqref{eq:psi1}, and \eqref{eq:psi2} so that the following limits hold:
		\begin{align}
		\varlimsup_{n\to \infty} \alpha_{1,n} \leq 0, \quad \varlimsup_{n\to \infty}\alpha_{2,n} \leq 0,
		\label{eq-definition-constant-constraints-typeI-errors}
		\end{align}
		\begin{IEEEeqnarray}{rCl}
		\theta_1 &\leq   \varliminf_{n \to \infty}-\frac{1}{n} \log \beta_{1,n}, \quad 
		\theta_2& \leq  \varliminf_{n \to \infty}-\frac{1}{n} \log \beta_{2,n},\label{eq:theta}
		\end{IEEEeqnarray}
		and 
		\begin{IEEEeqnarray}{rClCl}\label{eq-rate-constraints}
		\varlimsup_{n\to \infty} \frac{1}{n}\log \mathsf{W}_{1,n} &\leq R_1, \quad
		\varlimsup_{n\to \infty} \frac{1}{n}\log \mathsf{W}_{2,n}  &\leq R_2 .
		\label{eq-definition-achievable-rates}
		\end{IEEEeqnarray}
	\end{definition}

	\begin{definition}[Error-Exponents Region under Rate-Constraints] \label{def:error-exponent-region}
		For $\bar{h}_1\in\{0,1\}$ and  rates $R_1, R_2 \geq 0$ the  closure of the set of all achievable exponent pairs $(\theta_1, \theta_2)$ is called the \emph{error-exponents region} $\mathcal{E}(R_1, R_2)$. %For $\epsilon_1 \to 0$ and $\epsilon_2 \to 0$, the limit of the region $\mathcal{E}(R_1, R_2, \epsilon_1, \epsilon_2)$ will be denoted as $\mathcal{E}(R_1, R_2,0, 0)$, i.e., 
%		\begin{equation}
%		\mc E(R_1,R_2,0,0) = \bigcap_{\epsilon_1 > 0 , \epsilon_2>0 } \mc E(R_1,R_2,\epsilon_1,\epsilon_2) .
%		\end{equation}
	\end{definition}
	
	\bigskip

When both rates are zero, 
\begin{equation}
R_1=R_2=0,
\end{equation}
we are also interested in determining the exponents region with fixed communication alphabets of sizes:
\begin{subequations}\label{eq:fixed_trans}
\begin{IEEEeqnarray}{rCl}
	\mathsf{W}_{1,n} & = & \mathsf{W}_1 \geq 2 \\
		\mathsf{W}_{2,n}  &  =  & \mathsf{W}_2 \geq 2.
	\end{IEEEeqnarray}
\end{subequations}
\begin{definition}[Achievability with Fixed Communication Alphabets]%[Achievable Type-II Error Exponents under Constant Constraints on Type I Errors] 
	For $\bar{h}_1\in\{0,1\}$ and communication alphabet sizes $\W_1, W_2\geq 0$,
	an error-exponents pair $(\theta_1,\theta_2)$ is said achievable if for all blocklengths $n$ there exist  functions $\phi_{1,n}$, $\phi_{2,n}$, $\psi_{1,n}$ and $\psi_{2,n}$ as in \eqref{eq:phi1}, \eqref{eq:phi2}, \eqref{eq:psi1}, and \eqref{eq:psi2} so that \eqref{eq:fixed_trans} and the  limits in \eqref{eq-definition-constant-constraints-typeI-errors} and \eqref{eq:theta} hold.
%	\begin{align}
%	\varlimsup_{n\to \infty} \alpha_{1,n} \leq \epsilon, \quad \varlimsup_{n\to \infty}\alpha_{2,n} \leq \epsilon,
%	\label{eq-definition-constant-constraints-typeI-errors}
%	\end{align}
%	\begin{IEEEeqnarray}{rCl}
%		\theta_1 &\leq   \varliminf_{n \to \infty}-\frac{1}{n} \log \beta_{1,n}, \quad 
%		\theta_2& \leq  \varliminf_{n \to \infty}-\frac{1}{n} \log \beta_{2,n}.
%	\end{IEEEeqnarray}
%	and 
%	\begin{IEEEeqnarray}{rClCl}
%		\varlimsup_{n\to \infty} \frac{1}{n}\log \mathsf{W}_{1,n} &\leq R_1, \quad
%		\varlimsup_{n\to \infty} \frac{1}{n}\log \mathsf{W}_{1,n}  &\leq R_2 .
%		\label{eq-definition-achievable-rates}
%	\end{IEEEeqnarray}
\end{definition}

\begin{definition}[Error-Exponents Region for Fixed Communication Alphabets] \label{def:error-exponent-region-fixed-comm}
	For fixed $\bar{h}_1\in\{0,1\}$ and communication alphabet sizes $\W_1,\W_2\geq 0$, the  closure of the set of all achievable exponent pairs $(\theta_1, \theta_2)$ is called the \emph{error-exponents region} $\mathcal{E}_0(\W_1,\W_2)$. %For $\epsilon_1 \to 0$ and $\epsilon_2 \to 0$, the limit of the region $\mathcal{E}(R_1, R_2, \epsilon_1, \epsilon_2)$ will be denoted as $\mathcal{E}(R_1, R_2,0, 0)$, i.e., 
	%		\begin{equation}
	%		\mc E(R_1,R_2,0,0) = \bigcap_{\epsilon_1 > 0 , \epsilon_2>0 } \mc E(R_1,R_2,\epsilon_1,\epsilon_2) .
	%		\end{equation}
\end{definition}
	%So we use notation $\mc E(R_1,R_2)$ under the rate constraints \eqref{eq-rate-constraints} and notation $\mc E_0(\W_1,\W_2)$ under the more stringent fixed-alphabet constraints \eqref{eq:fixed_trans}.

%	The main interest  of this paper is to characterize  the error-exponents region. We will separately treat the case of \emph{coherent detection} where
%	\begin{equation}
%	i_1=i_2
%	\end{equation}
%	and \emph{concurrent detection} where
%	 	\begin{equation}
%	 	i_1\neq i_2.
%	 	\end{equation}
	
%\todo{	In our investigation of the role of cooperation we will also sometimes focus on the model without cooperation, i.e., the model of Figure~\ref{fig-system-model} but without the link of capacity $R_2$. For this model, the definition are similar to those of the model of Figure~\ref{fig-system-model}. For instance the encoding mapping at Sensor and the decision function at Decoder 1 are still given by \eqref{eq:phi1} and \eqref{eq:psi1} respectively; and the decision function at Decoder 2 as given by \eqref{eq:psi2} is replaced with 
%
%	\begin{equation}\label{eq:psi2-no-coop}
%	\psi_{2,n}\colon  \mc M_1  \times \mc Y^n_2 \rightarrow \{1,\ldots,\M\}, 
%	\end{equation}
%	for this scenario we define the \emph{error-exponents} region $\mathcal{E}_{\text{nc}}(R_1)$ similar to \ref{def:error-exponent region}.}
	
%------------------

\begin{remark}\label{rem:zhou-lai}
The model of Fig. 1  was also considered in \cite{ZL18} in a special case of testing against independence where
\begin{IEEEeqnarray}{rCl}\label{eq:zhou-lai}
 \bar{P}_{XY_1Y_2} = P_{XY_1} P_{Y_2}.
\end{IEEEeqnarray}
Moreover, in \cite{ZL18} only Detector 2 guesses the binary hypothesis but not Detector 1.
\end{remark}

\subsection{Some Degenerate Cases}

In some special cases,  the described setup degenerates and the error-exponents region is the same as  in a setup without cooperation or in a setup with a single centralized detector.

We first consider a setup where cooperation is not beneficial.

\begin{proposition}\label{prop:same2}
	Assume  the Markov chain $X\mkv Y_2 \mkv Y_1$  under both hypotheses with identical law $P_{Y_1|Y_2}=\bar{P}_{Y_1|Y_2}$:
	\begin{subequations}\label{eq:markov2}
		\begin{IEEEeqnarray}{rcl}
			P_{XY_1Y_2} &=& P_{XY_2}P_{Y_1|Y_2}\\
			\bar{P}_{XY_1Y_2} &=& \bar{P}_{XY_2}P_{Y_1|Y_2}.
		\end{IEEEeqnarray}
	\end{subequations} In this case, irrespective of the cooperation rate $R_2\geq 0$ and of the value of $\bar h_1\in\{0,1\}$, the exponent region $\mathcal{E}(R_1,R_2)$ and $\mathcal{E}_0(\W_1,\W_2)$ coincides with the exponent region of the  scenario without cooperation (see Fig.~\ref{fig:mkv-ch-equivalent}).
\begin{figure}[!ht]
	\begin{center}
		\includegraphics[width=.66\linewidth]{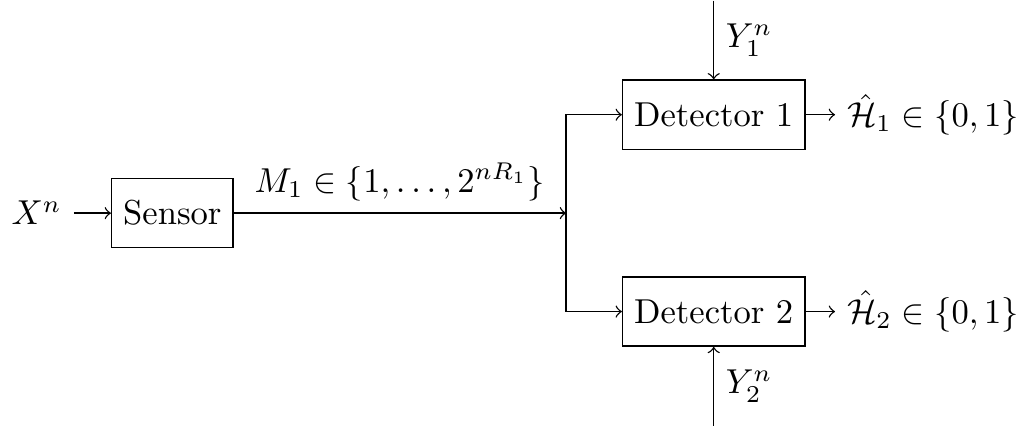}
		\caption{Equivalent system without cooperation when $X \mkv Y_2 \mkv Y_1$ under both hypotheses.}
		\label{fig:mkv-ch-equivalent}
	\end{center}
	\vspace{-5mm}
\end{figure}
\end{proposition}
	\begin{IEEEproof}
	The exponents regions $\mathcal{E}(R_1,R_2)$  and $\mathcal{E}_0(\W_1,\W_2)$ of the original setup cannot be larger than the exponent regions of an enhanced setup (with cooperation) where Detector 2 not only observes $Y_2^n$ but also $Y_1^n$. But in this new setup,  the cooperation link is  useless because Detector 2 can generate the cooperation message locally. Moreover, without cooperation, the observation $Y_1^n$ is not beneficial because the conditional laws $P_{Y_1|XY_2}$ and $\bar{P}_{Y_1|XY_2}$ coincide and only depend on $Y_2$, and so Detector 2 can generate a statistically equivalent observation to $Y_1^n$ itself based only on $Y_2^n$. %but not on $X$.  
	By these arguments,  the exponents regions $\mathcal{E}(R_1,R_2)$  and $\mathcal{E}_0(\W_1,\W_2)$ of the original setup are not larger than the one of the setup without cooperation. But $\mathcal{E}(R_1,R_2)$ and $\mathcal{E}_0(\W_1,\W_2)$ can also not be smaller than the exponents regions of the same setup but without cooperation, because the latter setup can be mimicked in the former. This concludes the proof.
\end{IEEEproof}

\bigskip

We now consider a setup that is equivalent to a setup with a single centralized detector.

\begin{figure}[!ht]
	\begin{center}
		\includegraphics[width=.8\linewidth]{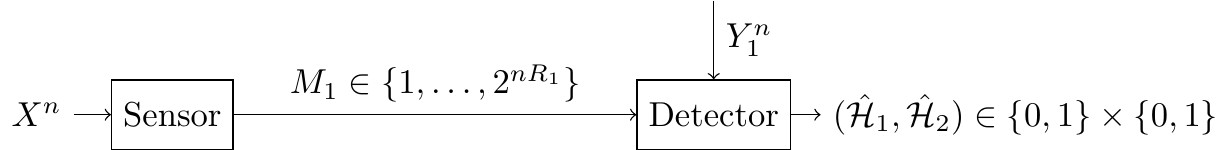}
		\caption{Equivalent point to point system when $X \mkv Y_1 \mkv Y_2$ under both hypotheses.} %{\color{red}Please make Fig 2. and FIg. 3 consistent in what you write on the cooperation link from the sensor to the detectors. J'ai corrigé et replacé dans le texte }}
		\label{fig:mkv-ch-equivalent-simple}
	\end{center}
	\vspace{-5mm}
\end{figure}
\begin{proposition}\label{prop:same}
	Assume the Markov chain $X \mkv Y_1 \mkv Y_2$ holds under both hypotheses with identical law $P_{Y_2|Y_1}=\bar{P}_{Y_2|Y_1}$. I.e.,
	\begin{subequations}\label{eq:markov}
		\begin{eqnarray}
		P_{XY_1Y_2} &= P_{XY_1}P_{Y_2|Y_1}\\
		\bar{P}_{XY_1Y_2} &= \bar{P}_{XY_1}P_{Y_2|Y_1}.
		\end{eqnarray}
	\end{subequations} In this case, irrespective of the cooperation rate $R_2\geq 0$ and of the value of $h_1\in\{0,1\}$, the exponent regions $\mathcal{E}(R_1,R_2)$ and $\mathcal{E}_0(\W_1,\W_2)$ coincide with the exponents regions of the  scenario in Fig.~\ref{fig:mkv-ch-equivalent} without cooperation and where both Detectors observe $Y_1^n$ but not $Y_2^n$.  As a consequence, they also coincide with the exponents regions of the scenario in Fig.~\ref{fig:mkv-ch-equivalent-simple} with  a single Detector observing $Y_1^n$ that takes both decisions $\hat{\mathcal{H}}_1$ and $\hat{\mathcal{H}}_2$.
\end{proposition}
 \begin{IEEEproof}
	The exponents regions $\mathcal{E}(R_1,R_2)$  and $\mathcal{E}_0(\W_1,\W_2)$ of the original setup cannot be larger than the exponents regions of an enhanced setup (with cooperation) where Detector 2 not only observes  $Y_2^n$ but also $Y_1^n$. Since Detector 2  can generate an observation that is  statistically equivalent to $Y_2^n$ given $Y_1^n$, the exponents regions are no larger than in the setup where both detectors observe   $Y_1^n$ but not $Y_2^n$. Furthermore, since allowing the two detectors to fully cooperate in their decision can only increase the exponents regions, the regions $\mathcal{E}(R_1,R_2)$ and $\mathcal{E}_0(\W_1,\W_2)$  must be included in the exponents regions of the setup in  Fig.~\ref{fig:mkv-ch-equivalent-simple} where a single detector takes both decisions.
	
%	In this latter setup the cooperation link is useless %because the conditional laws $P_{Y_2|XY_1}$ and $\bar{P}_{Y_2|XY_1}$ coincide and only depend on $Y_1$ but not on $X$ and so
%	since Detector 2 observes everything that Detector 1 observes and thus can generate the cooperation message locally. Therefore, the exponents regions $\mathcal{E}(R_1,R_2)$ and $\mathcal{E}_0(\W_1,\W_2)$ are not larger than in a  scenario without cooperation where both re Fig.~\ref{fig:mkv-ch-equivalent-simple} where  Detectors $2$ observes $Y_1^n$ but not $Y_2^n$.
	
	On the other hand, $\mathcal{E}(R_1,R_2)$ and $\mathcal{E}_0(\W_1,\W_2)$ can also not be smaller than the exponents regions of the setup in Fig.~\ref{fig:mkv-ch-equivalent-simple}. In fact, in the original setup, Detector $1$ can mimick the single central detector and forward the   decision $\hat{H}_2$ to Detector $2$, which follows this decision. This strategy requires only a single cooperation bit and can thus be implemented  irrespective to the available cooperation rate $R_2\geq 0$. %. But the setup of Fig.~\ref{fig:mkv-ch-equivalent-simple} is equivalent to the scenario of Fig.~\ref{fig:mkv-ch-equivalent} where Detector $2$ observes $Y_1^n$ and not $Y_2^n$.
	 This conclude the proof.
\end{IEEEproof}

\section{Results for Fixed Communication Alphabets}\label{secIII} %Zero-Rate \texorpdfstring{$R_1=R_2=0$}{R1 = R2 =0}}

We start by presenting our results for the fixed-alphabets case, so we assume \eqref{eq:fixed_trans} and are interested in the error-exponents region $\mathcal{E}_0(\mathsf{W}_1,\mathsf{W}_2)$. For  simplicity, we  assume that %$P_{Y_2}^{(1)}$ differs from $P_{Y_2}^{(2)}$  and moreover 
$P_{XY_1}(x,y_1) > 0 $ and $\bar P_{XY_1Y_2}(x,y_1,y_2) > 0$ for all $(x,y_1,y_2)\in\mathcal{X}_1\times \mathcal{Y}_2\times \mathcal{Y}_2$.

Our main finding in this section is the exact characterization of the optimal exponents region $\mathcal{E}_0(\mathsf{W}_1,\mathsf{W}_2)$ for all possible parameters. %Our results in particular show that the optimal exponents region shows a tradeoff between the two exponents $\theta_1$ and $\theta_2$ only for concurrent detection $\bar{h}_1=0$ with $P_X \neq \bar{P}_X$ and $\W_1=\W_2=2$. In all other cases, the optimal exponents region $\mathcal{E}_0(\mathsf{W}_1,\mathsf{W}_2)$  is a \emph{rectangle} and there is  no such tradeoff.  In fact, it is not difficult to see that in these latter cases, each terminal can simultaneously send the informations desired by each of the decision  centers. This explains the absence of a tradeoff. In the exceptional case where $\bar{h}_1=0$ with $P_X \neq \bar{P}_X$ and $\W_1=\W_2=2$, the alphabet sizes are not sufficient to simultaneously convey to each decision center its desired information.  
	
% $\mathcal{E}_0(\mathsf{W}_1,\mathsf{W}_2)$ shows one  are two different for coherent detection, the error-exponents region $\mathcal{E}_0(\mathsf{W}_1,\mathsf{W}_2)$ is a straightforward extension of the result in 
%\cite{SP92}. For non-coherent detection there is a twist. 
\subsection{Coherent Detection and Concurrent Detection with  $P_X = \bar{P}_X$}
In this sense, the following Propositions~\ref{prop-coherent}--\ref{prop-concurrent-no-depletion} are rather straightforward and we omit most of their proofs.  Proposition~\ref{prop-concurrent} is the main result of this section. 

\begin{proposition}[Coherent Detection]\label{prop-coherent}
	For coherent detection, $\bar{h}_1=1$, and for all values $\mathsf{W}_1 \geq 2$ and $\mathsf{W}_2 \geq 2$, the exponents region $\mathcal{E}_0(\mathsf{W}_1,\mathsf{W}_2)$ is the set of all nonnegative rate pairs $(\theta_1,\theta_2)$  satisfying
	\begin{IEEEeqnarray}{rCl}
		\theta_1& \leq& \min_{\begin{subarray}{c}\tilde P_{XY_1}\colon \;\tilde P_{X} = {P}_X \\ \tilde P_{Y_1} = {P}_{Y_1}  \end{subarray}} D\Big(\tilde P_{XY_1}\|\bar P_{XY_1}\Big) \label{eq:converse1}\\
		\theta_2& \leq &\min_{\begin{subarray}{c}\tilde P_{XY_1Y_2}\colon \; \tilde P_{X} = {P}_X \\ \tilde P_{Y_1} = {P}_{Y_1}, \; \tilde P_{Y_2} = {P}_{Y_2} \end{subarray}} D\Big(\tilde P_{XY_1Y_2}\|\bar{P}_{XY_1Y_2}\Big).\label{eq:converse2}
	\end{IEEEeqnarray}
\end{proposition}
\begin{IEEEproof}
	The achievability follows by a  similar scheme as in \cite{H87} where the sensor and Detector~$1$ check whether their observed sequences are typical or not according to the P-distribution and send the message 1 if so, and 0 if not. The detectors declare $\mathcal{H}=0$ if all received messages are ones and their observed sequence is typical. Otherwise, they declare  $\mathcal{H}=1$.
	The converse to \eqref{eq:converse1} follows directly from \cite{SP92}. The converse to \eqref{eq:converse2} is similar and proved for completeness in Appendix~\ref{app:converse2}. 
\end{IEEEproof}

\begin{proposition}[Concurrent Detection with $P_X=\bar{P}_X$]\label{prop-concurrent}
	Under concurrent detection, i.e., $\bar{h}_1=0$, and  when $P_X=\bar{P}_X$, then for all values $\mathsf{W}_1 \geq 2$ and $\mathsf{W}_2 \geq 2$, the exponents region $\mathcal{E}_0(\mathsf{W}_1,\mathsf{W}_2)$ is the set of all nonnegative rate pairs $(\theta_1,\theta_2)$  satisfying
	\begin{IEEEeqnarray}{rCl}
		\theta_1& \leq& \min_{\begin{subarray}{c}\tilde P_{XY_1}\colon \;  \tilde P_{X} = {P}_X \\ \tilde P_{Y_1} = \bar{P}_{Y_1}  \end{subarray}} D\Big(\tilde P_{XY_1}\| P_{XY_1}\Big) \label{eq:converse1-n}\\
		\theta_2& \leq &\min_{\begin{subarray}{c}\tilde P_{XY_1Y_2}\colon \; \tilde P_{X} =  {P}_X \\ \tilde P_{Y_1} =  {P}_{Y_1}, \; \tilde P_{Y_2} =  {P}_{Y_2} \end{subarray}} D\Big(\tilde P_{XY_1Y_2}\|\bar{P}_{XY_1Y_2}\Big).\label{eq:converse2-c}
	\end{IEEEeqnarray}
\end{proposition}
\begin{IEEEproof}The achievability follows by a slight generalization of the  previous scheme. The idea is that Detector  1 decides on $\mathcal{H}=1$ only if $X^n$ and $Y_1^n$ are typical according to the $\bar{P}$ distributions. Detector 2 decides on $\mathcal{H}=0$ only if $X^n$, $Y_1^n$, and $Y_2^n$ are typical according to the $P$ distributions. Since $P_X=\bar{P}_X$ this information can be conveyed to both detectors whenever $\W_1,\W_2\geq 2$. The converse is similar to the one for Proposition~\ref{prop-coherent} and omitted.
\end{IEEEproof}

\subsection{Concurrent Detection with \texorpdfstring{$P_X \neq\bar{P}_X$}{PX != PXB}}\label{sec:cocncurent-zerorate}

We now consider concurrent detection, $\bar{h}_1=0$, and  $P_X \neq \bar{P}_X$. Here the optimal exponents region depends on whether the alphabet size $\W_1$ equals $2$ or is larger.  %eqref{eq-concurrent} holds, but not \eqref{eq-minimum-rate}.
 We first assume
\begin{equation}
\W_1\geq 3 \qquad \textnormal{and} \qquad \W_2 \geq 2,
\end{equation}
and present a coding scheme for this scenario. Pick a small positive number $\mu>0$ such that the typical sets $\mc T_{\mu}^n(P_X)$ and $\mc T_{\mu}^n(\bar P_X)$ do not intersect:
\begin{equation}\label{eq:mu}
\mathcal{T}^n _{\mu}(P_X)\cap  \mathcal{T}^n_{\mu}(\bar{P}_X)= \emptyset.
\end{equation}
\underline{\textit{Sensor:}} Given that it observes $X^n=x^n$, it sends 
\begin{equation}
M_1 = \left\{ \begin{matrix}	0 & \textnormal{ if } x^n \in \mathcal{T}_{\mu}^n(P_X) \\
1	& \textnormal{ if } x^n \in \mathcal{T}_{\mu}^n(\bar{P}_X) \\
2 & \textnormal{ otherwise}.
\end{matrix}\right.
\end{equation}

\underline{\textit{Detector 1:}} Given that it observes $Y_1^n=y_1^n$ and $M_1=m_1$, it decides  
\begin{equation}
\hat{\mathcal{H}}_1 = \left\{ \begin{matrix}	1 & \textnormal{ if } m_1=1 \quad \textnormal{and} \quad y_1^n \in \mathcal{T}_{\mu}^n(\bar{P}_{Y_1}) \\
0	 & \textnormal{ otherwise}.
\end{matrix}\right.
\end{equation}
It sends 
\begin{equation}
M_2 = \left\{ \begin{matrix}	0 & \textnormal{ if } m_1=0 \quad \textnormal{and} \quad  y_1^n \in \mathcal{T}_{\mu}^n(P_{Y_1}) \\
1	&\textnormal{ otherwise}
\end{matrix}\right.
\end{equation} to Detector 2.

\underline{\textit{Detector 2:}} %\underline{\textit{Detector 2:}}
 Given that it observes $Y_2^n=y_2^n$ and messages $M_1=m_1$ and $M_2=m_2$, it decides  
\begin{equation}\label{eq:coh_det2}
\hat{\mathcal{H}}_2 = \left\{ \begin{matrix}	0 & \textnormal{ if  } m_1=m_2=0 \quad \textnormal{and} \quad  y_2^n \in \mathcal{T}_{\mu}^n(P_{Y_2}) \\
1	 & \textnormal{ otherwise}.
\end{matrix}\right.
\end{equation}%As for  coherent detection, see \eqref{eq:coh_det2}. %Given that it observes $Y_2^n=y_2^n$ and messages $M_1=m_1$ and $M_2=m_2$, it decides  
%\begin{equation}
%\hat{\mathcal{H}}_2 = \left\{ \begin{matrix}	0 & \textnormal{ if  } m_1=m_2=0 \quad \textnormal{and} \quad  y_2^n \in \mathcal{T}_{\mu}^n(P_{Y_2}) \\
%1	 & \textnormal{ otherwise}.
%\end{matrix}\right.
%\end{equation}

\begin{proposition}[Concurrent Dection, $P_X\neq \bar{P}_X$ and $\W_1\geq 3$] \label{prop-concurrent-no-depletion}
	Under concurrent detection and for all values $\mathsf{W}_1 \geq 3$ and $\mathsf{W}_2 \geq 2$, the exponents region $\mathcal{E}_0(\mathsf{W}_1,\mathsf{W}_2)$ is the set of all nonnegative rate pairs $(\theta_1,\theta_2)$  satisfying
	\begin{IEEEeqnarray}{rCl}
		\theta_1  &\leq & \min_{\begin{subarray}{c}\tilde P_{XY_1}\colon \; \tilde P_{X} =\bar {P}_X \\ \tilde P_{Y_1} = \bar {P}_{Y_1}  \end{subarray}} D\Big(\tilde P_{XY_1}\| P_{XY_1}\Big)\\
		\theta_2 &\leq& \min_{\begin{subarray}{c}\tilde P_{XY_1Y_2}\colon \; \tilde P_{X} = {P}_X \\ \tilde P_{Y_1} = {P}_{Y_1}, \; \tilde P_{Y_2} = {P}_{Y_2} \end{subarray}} D\Big(\tilde P_{XY_1Y_2}\|\bar{P}_{XY_1Y_2}\Big).\label{eq:theta2}
	\end{IEEEeqnarray}
\end{proposition}
\begin{IEEEproof}
	The achievability follows by the above coding scheme; and the converse is similar to that of Proposition~\ref{prop-coherent}.
\end{IEEEproof}

The exponents region $\mathcal{E}_0(\mathsf{W}_1,\mathsf{W}_2)$ in these first three Propositions \ref{prop-coherent}--\ref{prop-concurrent-no-depletion} is rectangular, and each of  the detectors  can simultaneouly achieve the optimal exponent as if it were the only detector in the system. As we see in the following, this is not  always the case.

In the rest of this section, we assume 
\begin{equation}
\mathsf{W}_1 =2 \quad \textnormal{and} \quad \mathsf{W}_2\geq 2,
\end{equation}
and present the optimal exponents region for this case. It is achieved by the following coding scheme. 

Pick a real number $r$, a small positive number $\mu>0$ satisfying \eqref{eq:mu}, and the function $b\colon \{0,1\} \to \{0,1\}$ either as 
\begin{equation}\label{eq:same}
b(0)=b(1)=0
\end{equation}
or as 
\begin{equation}\label{eq:different}
b(0)=0 \qquad \textnormal{and} \qquad b(1)=1.
\end{equation}

We then assign each type $\pi \in \mathcal{P}(\mathcal{X}^n)$ that satisfies
\begin{equation}
| \pi - P_X | > \mu \qquad \textnormal{and}\qquad | \pi -\bar{P}_x| >\mu
\end{equation} to one of two sets $\Gamma_0$ or $\Gamma_1$.  If $b(0)=b(1)=0$, then we assign all these types to the set $\Gamma_1$. Otherwise, we assign  them between the two sets according to the following rule:
\begin{equation}\label{rule-partition-creation}
\pi \in \Gamma_{b(1)} \quad \Longleftrightarrow \quad \min_{\begin{subarray}\tilde P_{XY_1}:\\ \tilde P_{X} = \pi \\ \tilde P_{Y_1} = \bar{P}_{Y_1}  \end{subarray}} D\Big(\tilde P_{XY_1}\|{P}_{XY_1}\Big) +r \geq \min_{\begin{subarray}\tilde P_{XY_1Y_2}:\\ \tilde P_{X} = \pi \\ \tilde P_{Y_1} = {P}_{Y_1} \\  \tilde P_{Y_2} = {P}_{Y_2} \end{subarray}}  D\Big(\tilde P_{XY_1Y_2}\|\bar P_{XY_1Y_2}\Big),
\end{equation}
and $\pi \in \Gamma_{b(0)}$ otherwise.
Given that it observes $X^n=x^n$, it sends
\begin{equation}
M_1= \left\{ \begin{matrix}	b(0) & \textnormal{ if } x^n \in \mathcal{T}_{\mu}^n(P_X) \\
b(1)	& \textnormal{ if } x^n \in \mathcal{T}_{\mu}^n(\bar{P}_X) \\
0 &  \textnormal{ if } \textnormal{tp}(x^n) \in \Gamma_0 \\
1&  \textnormal{ if }\textnormal{tp}(x^n) \in \Gamma_1.
\end{matrix}\right.
\end{equation}

\underline{\textit{Detector 1:}} Given that it observes $Y_1^n=y_1^n$ and $M_1=m_1$, it decides  
\begin{equation}
\hat{\mathcal{H}}_1 = \left\{ \begin{matrix}	1 & \textnormal{ if } m_1=b(1) \quad \textnormal{and} \quad y_1^n \in \mathcal{T}_{\mu}^n(\bar{P}_{Y_1}) \\
0	 & \textnormal{ otherwise}.
\end{matrix}\right.
\end{equation}
It sends 
\begin{equation}
M_2 = \left\{ \begin{matrix}	0 & \textnormal{ if } m_1=b(0) \quad \textnormal{and} \quad  y_1^n \in \mathcal{T}_{\mu}^n(P_{Y_1}) \\
1	&\textnormal{ otherwise}
\end{matrix}\right.
\end{equation} to Detector 2.

\underline{\textit{Detector 2:}} Given that it observes $Y_2^n=y_2^n$ and messages $M_1=m_1$ and $M_2=m_2$, it decides  
\begin{equation}
\hat{\mathcal{H}}_2 = \left\{ \begin{matrix}	0 & \textnormal{ if  } m_1=b(0)\quad \textnormal{and} \quad m_2=0 \quad \textnormal{and} \quad  y_2^n \in \mathcal{T}_{\mu}^n(P_{Y_2}) \\
1	 & \textnormal{ otherwise}.
\end{matrix}\right.
\end{equation}

The described scheme achieves the following optimal error-exponents region.

\begin{theorem}[Concurrent Detection, $P_X\neq \bar{P}_X$, and $\W_1=2$] \label{theorem: optimal concurrent hypothesis testing zero rate with cooperation and depletion at the encoder}Under concurrent detection and for all values $\mathsf{W}_1 = 2$ and $\mathsf{W}_2 \geq 2$, the exponents region $\mathcal{E}_0(\mathsf{W}_1,\mathsf{W}_2)$ is the set of all nonnegative rate pairs $(\theta_1,\theta_2)$ that satisfy
	\begin{IEEEeqnarray}{rCl}
		\theta_1 &\leq&  
		\min_{\substack{\tilde P_{XY_1}: \\[.2ex] \tilde P_{X} \in \Gamma_{b(1)} \\[.2ex] %=\tilde{P}_X\\  
				\tilde P_{Y_1}=\bar{P}_{Y_1} }}% \\ (\tilde{P}_X,m) \in \bigcup_{i=1}^\mathsf{W}{\phi_{i,X} \times \mc B_i \backslash \{k\} } \end{array}}
		\!\!\!{D\Big(\tilde P_{XY_1}\|P_{XY_1}\Big)},  \\
		\theta_2 &\leq&   \min_{\substack{\tilde P_{XY_1Y_2}\colon \\ \tilde P_{X} \in \Gamma_{b(0)} ,\\  \tilde P_{Y_1}={P}_{Y_1},\; 
				\tilde P_{Y_2}={P}_{Y_2} }}
		{D\Big(\tilde P_{XY_1Y_2}\|\bar P_{XY_1Y_2}\Big)}.
	\end{IEEEeqnarray}
	for some real $r$ and one of the mappings in \eqref{eq:same} and \eqref{eq:different}, and the corresponding sets $\Gamma_0$ and $\Gamma_1$.
\end{theorem}
\begin{IEEEproof}
		The achievability follows by the described scheme, by Sanov's theorem, and  by noting that $\hat{\mathcal{H}}_1=1$ iff,
		\begin{IEEEeqnarray}{r}	
		\textnormal{tp}(x^n) \in  \Gamma_{b(1)} \qquad \textnormal{and} \qquad 	y_1^n \in \mathcal{T}_{\mu}^n({P}_{Y_1}) 
		\end{IEEEeqnarray} whereas $\hat{\mathcal{H}}_2=0$, if, and only if, 
		\begin{IEEEeqnarray}{r}	
	\textnormal{tp}(x^n) \in  \Gamma_{b(0)}\quad \textnormal{and} \quad 	y_1^n \in \mathcal{T}_{\mu}^n({P}_{Y_1}) \quad \textnormal{and} \quad 	y_2^n \in \mathcal{T}_{\mu}^n({P}_{Y_2}) .
		\end{IEEEeqnarray}
		The converse is proved in Appendix~\ref{App-B}.
\end{IEEEproof} 
\bigskip

\begin{remark}[Sending a sublinear number of bits]
	A close inspection of the converse proofs for Propositions~\ref{prop-coherent}--\ref{prop-concurrent-no-depletion} %and Theorem~\ref{theorem: optimal concurrent hypothesis testing zero rate with cooperation and depletion at the encoder} 
	shows that they remain valid when the alphabet sizes are not fixed but grow sublinearly in the blocklength~$n$, i.e., when
	\begin{equation}
	\varlimsup_{n\to \infty } \frac{\W_{i,n}}{n} =0, \qquad i\in\{1,2\}.
	\end{equation}
\end{remark}
\bigskip
\begin{remark}[Extension to many hypotheses]
Most of the  results in this section can easily be extended to a scenario with more than two hypotheses.
%In particular, with $\mathsf{H}\geq 3$ hypotheses, the optimal exponents region  $\mathcal{E}_{0}(\W_1,\W_2)$ continues to be a rectangle (without any tradeoff between the two detectors) under coherent detection and under concurrent detection, e.g., when $\W_2 \geq \mathsf{H}$ and the number of different  marginal distributions of $X^n$ does not exceed $\mathsf{H}-\W_1$.
For $\mathsf{H}=2$ the exponents region  showed a tradeoff in the exponents under concurrent detection only when $\W_1=\W_2=2$.
In contrast, for $\mathsf{H}\geq 3$, a tradeoff arises for a variety of pairs $\W_1,\W_2$. In general, the minimum required values for $\W_1$ and $\W_2$ leading to a rectangular exponents region coincides respectively with the number of hypotheses which have distinct $X$-marginals and the number of hypothesis which have distinct $Y_1$-marginals. 
% So for large number of hypotheses there can be a tradeoff in the exponents region even when the number of bits to be sent is large. 

%As remarked, the exponents region $\mathcal{E}_{0}(\W_1,\W_2)$ only shows a tradeoff between the two exponents under concurrent detection and when $
\end{remark}

\subsection{Benefits of Cooperation}

To discuss the benefits of cooperation, we quickly state the optimal exponents region without cooperation,  i.e., for\footnote{Equivalently, the no cooperation setup could be parametrized as $\W_2=1$.}
\begin{equation}
\W_2=0. 
\end{equation}
They were determined in \cite{EWZ18}. Under coherent detection or under concurrent detection with $P_X =\bar{P}_X$ or $\W_1\geq 3$, the exponents region $\mathcal{E}_{0}(\W_1,\W_2=0)$ are similar to Propositions \ref{prop-coherent}--\ref{prop-concurrent-no-depletion} but with a modified constraint on $\theta_2$. More precisely,  Propositions \ref{prop-coherent}--\ref{prop-concurrent-no-depletion} remain valid for $\W_2=0$ if the constraints on 
$\theta_2$, \eqref{eq:converse2}, \eqref{eq:converse2-c}, \eqref{eq:theta2} are replaced by
\begin{equation}
		\theta_2 \leq   \min_{\substack{\tilde P_{XY_2}\colon \\ \tilde P_{X} P_X,\\   
				\tilde P_{Y_2}={P}_{Y_2} }}
		{D\Big(\tilde P_{XY_2}\|\bar P_{XY_2}\Big)}.\nonumber
\end{equation}
So, in these  scenarios,
the exponents region is  a rectangle both in the case with and without cooperation, and with cooperation the $\theta_2$-side of the rectangle is increased 
by the  quantity
 \begin{equation}
 \min_{\begin{subarray}{c}\tilde P_{XY_1Y_2}\colon \; \tilde P_{X} = {P}_X \\ \tilde P_{Y_1} = {P}_{Y_1}, \; \tilde P_{Y_2} = {P}_{Y_2} \end{subarray}} D\Big(\tilde P_{XY_1Y_2}\|\bar{P}_{XY_1Y_2}\Big) - \min_{\begin{subarray}{c}\tilde P_{XY_1Y_2}\colon \;\\ \tilde P_{X} = {P}_X ; \\ \tilde P_{Y_2} = {P}_{Y_2} \end{subarray}} D\Big(\tilde P_{XY_2}\|\bar{P}_{XY_2}\Big).\label{eq:benefit}
 \end{equation}
	
%	It is not hard to see that the  quantity in \eqref{eq:benefit} is  unbounded. In fact, consider binary sources where $\bar{P}_1$  is deterministic but $P_{Y_1}$ is not. In this case, the first term 
%	\begin{equation} \min_{\begin{subarray}\tilde P_{XY_1Y_2}\colon \; \tilde P_{X} = {P}_X \\ \tilde P_{Y_1} = {P}_{Y_1}, \; \tilde P_{Y_2} = {P}_{Y_2} \end{subarray}} D\Big(\tilde P_{XY_1Y_2}\|\bar{P}_{XY_1Y_2}\Big) =\infty,
%	\end{equation}  whereas the second term can be finite: 
%	\begin{equation}\min_{\begin{subarray}{c}\tilde P_{XY_1Y_2}\colon \; \tilde P_{X} = {P}_X ; \\ \qquad \quad \tilde P_{Y_2} = {P}_{Y_2} \end{subarray}} D\Big(\tilde P_{XY_2}\|\bar{P}_{XY_2}\Big) < \infty.
%		\end{equation}  
	
Under concurrent detection  when $P_X \neq \bar{P}_X$ and $\W_1=2$, the exponents region is not a rectangle, but there is a tradeoff between the two exponents. In this case, it seems  difficult to  quantify the cooperation benefit in general.
% It can  again be unbounded, in particular in the simple example given above. 

%\subsection{Numerical Example}
We  now  present an example for concurrent detection with $P_X \neq \bar{P}_X$.   
\begin{example} \label{ex6}Consider a  setup where $X, Y_1, Y_2$ are binary with pmfs
	\begin{align}
	\left\{\begin{array}{cccc} P_{X Y_1 Y_2}(0,0,0) = 0.1990112  & P_{X Y_1 Y_2}(0,0,1) =0.16298342 \\ P_{X Y_1 Y_2}(0,1,0) =0.03853585 & P_{X Y_1 Y_2}(1,1) =0.0799361 \\ P_{X Y_1 Y_2}(1,0,0) = 0.09498084  & P_{X Y_1 Y_2}(1,0,1) = 0.03821415 \\ P_{X Y_1 Y_2}(1,1,0) =0.11018678 & P_{X Y_1 Y_2}(1,1,1) =0.08852474 \end{array} \right. \nonumber \\
	\left\{\begin{array}{cccc} \bar P_{X Y_1 Y_2}(0,0,0) = 0.19121486  & \bar P_{X Y_1 Y_2}(0,0,1) =0.12692116 \\ \bar P_{X Y_1 Y_2}(0,1,0) =0.19984744 & \bar P_{X Y_1 Y_2}(0,1,1) =0.1560087 \\ \bar P_{X Y_1 Y_2}(1,0,0) = 0.11718922  & \bar P_{X Y_1 Y_2}(1,0,1) = 0.19433398 \\ \bar P_{X Y_1 Y_2}(1,1,0) =0.04903381 & \bar P_{X Y_1 Y_2}(1,1,1) =0.15307775 \end{array} \right. \nonumber 
	\end{align}
\begin{figure}
	\begin{minipage}[t]{.49\textwidth}
			\includegraphics[width=7cm]{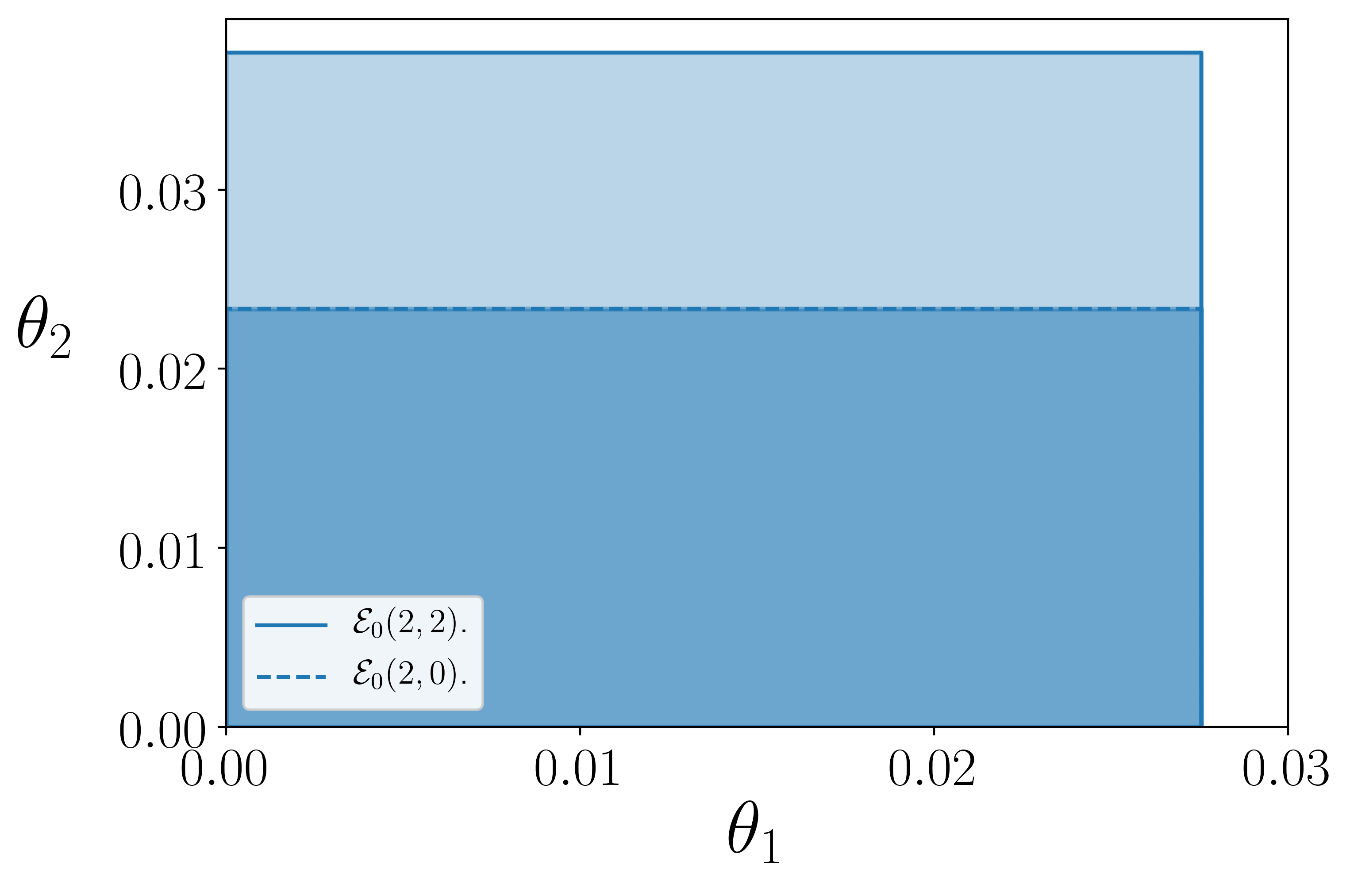}\vspace{-0.5cm}
	\end{minipage}
	\begin{minipage}[t]{.49\textwidth}
		\includegraphics[width=7cm]{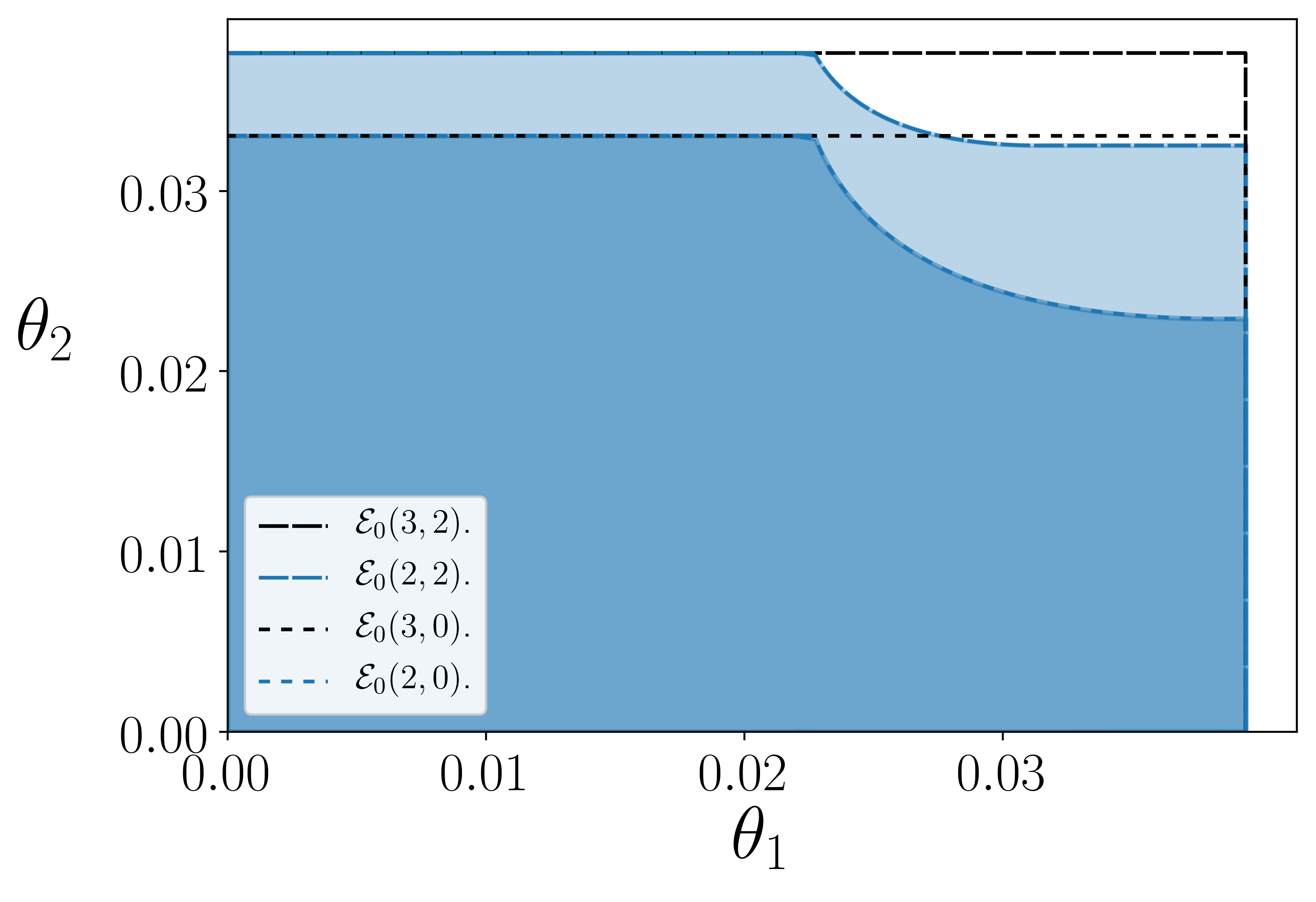}\vspace{-0.5cm}
\end{minipage}
\caption{Exponents region of Example~\ref{ex6}, see \cite{EZW19C} for implementation details. On the left: exponent regions $\mathcal{E}_0(2,2)$ and  $\mathcal{E}_0(2,0)$ for  coherent detection. On the right: exponent regions $\mathcal{E}_0(3,2)$,  $\mathcal{E}_0(3,0)$,  $\mathcal{E}_0(2,2)$, and  $\mathcal{E}_0(2,0)$ for  concurrent detection.}
%{\color{red} Please increase the font on the axis of the figures! => done}
\label{fig-opposite-exponent-region-2}
\end{figure}	

Fig.~\ref{fig-opposite-exponent-region-2} illustrates the exponents region for coherent and concurrent detection of Propositions~\ref{prop-coherent} and~\ref{prop-concurrent-no-depletion} and of Theorem~\ref{theorem: optimal concurrent hypothesis testing zero rate with cooperation and depletion at the encoder}. Specifically, the figure on the left shows the exponents region  with or without cooperation under coherent detection, and the figure on the right shows these exponents regions under concurrent detection. Under concurrent detection the exponents regions for $\W_1=3$ and $\W_1=2$ are shown.
%: $\mc E_0(3,2)$ and  $\mc E_0(3,0)$ are depicted by the dashed-dotted black and red rectangles and  $\mc E_0(2,2)$  and $\mc E_0(2,0)$ by the light and dark blue shaded regions. %The figure also shows the equivalent regions   without cooperation derived in \cite[Theorem~2]{EWZ18} in form of a dark blue area and a red rectangle. We can thus identify the gain of the cooperation link from Detector 1 to Detector 2. Notice that here it suffices to send a single cooperation bit.

\end{example}

%---------------------------------------------------------------

%\section{Rate Constraints: Extreme Cases and Previous Results}

\section{High-Rates Regime}\label{sec:large}
We now consider the other (trivial) extreme case where both links are of high rates so that under hypothesis $\mathcal{H}=h_1$, Detector 1 can learn the sequence $X^n$ with high probability and under $\mathcal{H}=0$, Detector 2 can learn both sequences $X^n$ and $Y_1^n$ with high probability. We will see that in this case both Detecor~1 attains the exponent of a centralized setup where it observes $(X^n,Y_1^n)$ and Detector~2 attains the exponent of a centralized setup where it observes $(X^n,Y_1^n,Y_2^n)$. 

We first consider coherent detection where $\bar{h}_1=1$. Pick a small $\epsilon$. The Sensor describes the sequence $X^n$ to both detectors if  $X^n \in \mathcal{T}_{\mu}^{(n)}(P_X)$, and otherwise it sends $0$. Detector~1 describes the sequence  $Y_1^n$  to Detector~2 if $(X^n,Y_1^n) \in \mathcal{T}_{\mu}^{(n)}(P_{XY_1})$, and otherwise it sends 0. The described coding scheme requires rates 
		\begin{IEEEeqnarray}{rCl}
			R_1 & \geq & H(X)+ \epsilon \\ 
			R_2 &\geq & H(Y_1|X) +\epsilon.
		\end{IEEEeqnarray}
Detector~1 decides on $\hat{\mathcal{H}}_1=1$, if the Sensor sent $0$ or itself it sent $0$. Otherwise it decides on $\hat{\mathcal{H}}_1=1$.
Detector~2 decides on $\hat{\mathcal{H}}_2=1$, if the Sensor or Detector~1 sent $0$. Otherwise it decides on $\hat{\mathcal{H}}_1=1$ if and only if its own observation $Y_2^n$ and the received sequences $X^n$ and $Y_1^n$ are jointly typical, $(X^n, Y_1^n, Y_2^n) \in \mathcal{T}_{\mu}^{(n)}(P_{XY_1Y_2})$.

The described scheme achieves the set of all nonnegative pairs $(\theta_1,\theta_2)$ satisfying
		\begin{subequations}
			\begin{align}
			\theta_1 &\leq  D(P_{XY_1}\|\bar{P}_{XY_1})\label{eq:cons1} \\
			\theta_2 &\leq D(P_{XY_1Y_2}\|\bar{P}_{XY_1Y_2}).\label{eq:cons2}
			\end{align}
			\label{optimal-exponents-centralized-case}
		\end{subequations}
		This set coincides with the optimal exponents region $\mathcal{E}(R_1,R_2)$, because it also coincides with the exponent region of a centralized setup where Detector 1 observes both $X^n$ and $Y_1^n$ and Detector 2 observes all $X^n$, $Y_1^n$, and $Y_2^n$.
		
		Consider now concurrent detection where $\bar{h}_1=0$. In this case, the Sensor  describes the sequence $X^n$ to both Detectors if  $X^n \in \mathcal{T}_{\mu}^{(n)}(P_X)$ or if $X^n \in \mathcal{T}_{\mu}^{(n)}(\bar{P}_X)$. Otherwise it sends $0$. Detector 1 describes the sequence  $Y_1^n$  to Detector 2 if $(X^n,Y_1^n) \in \mathcal{T}_{\mu}^{(n)}(P_{XY_1})$, and otherwise it sends 0. Detector 2 decides as above and Detector 1 decides on $\hat{\mathcal{H}}_1=0$ if and only if its own observation $Y_1^n$ and the described sequence $X^n$ are jointly typical, i.e., $(X^n Y_1^n) \in \mathcal{T}_\mu^n(\bar{P}_{XY_1})$. The coding scheme requires rates 
				\begin{IEEEeqnarray}{rCl}
					R_1 & \geq & \max\{ H(\bar{X}), H(X)\}+ \epsilon \\ 
					R_2 &\geq & H(Y_1|X) +  \epsilon.
				\end{IEEEeqnarray}
and achieves the set of all nonnegative pairs $(\theta_1,\theta_2)$ satisfying
		\begin{subequations}
			\begin{align}
			\theta_1 &\leq  D(\bar{P}_{XY_1}\|{P}_{XY_1})\label{eq:cons1-b} \\
			\theta_2 &\leq D(P_{XY_1Y_2}\|\bar{P}_{XY_1Y_2}).\label{eq:cons2-b}
			\end{align}
				\end{subequations}	
				Again, this set coincides with the optimal exponents region $\mathcal{E}(R_1,R_2)$ because it also coincides with the optimal exponents region when Detector 1 observes the pair $X^n,Y_1^n$ and Detector 2 observes $X^n$, $Y_1^n$, and $Y_2^n$.
%		\end{corollary}\begin{IEEEproof} The converse follows trivially by noting that the described region equals the set of all achievable exponents in a centralized setup where Detector 1 observes both $X^n$ and $Y_1^n$ and Detector 2 observes all three sequences $X^n,Y_1^n, Y_2^n$. Achievability follows from Theorem~\ref{theorem-lower-bounds-power-exponents-general-hypotheses-positive-rates} 
%		by setting $U=X$ and $V=Y_1$. 
%	\end{IEEEproof}

Both results remain valid without cooperation if the term $D(P_{XY_1Y_2} \| \bar{P}_{XY_1Y_2})$ limiting the second exponent $\theta_2$ is replaced by $D(P_{XY_2} \| \bar{P}_{XY_2})$. The benefit of cooperation is thus equal to $\mathbb{E}_{P_{XY_2}} [D(P_{Y_1|XY_2} \| \bar{P}_{Y_1|XY_2})]$ in both cases. 
%-----------------------------------------------------------------------------------------------------

\section{Results for Positive Rates \texorpdfstring{$R_1>0$}{R1>0}}\label{sec:moderate}

In this section, we assume that
\begin{equation}
R_1 >0.
\end{equation} 
The cooperation rate $R_2$ can be $0$ or larger.

		\subsection{Testing Against Independence under Coherent Detection}	
		We start with the special case of ``testing-against-independence" scenario under coherent detection, $\bar h_1=1$, where 
		\begin{IEEEeqnarray}{rCl}\label{eq: condition case 2}
			P_{XY_1Y_2} &= P_{X|Y_1Y_2}P_{Y_1} P_{Y_2} \\  \bar P_{XY_1Y_2} &= P_XP_{Y_1}P_{Y_2}.
		\end{IEEEeqnarray}
		Notice that this setup differs from the testing-against independence scenario in \cite{ZL18} (see also Remark~\ref{rem:zhou-lai}). 
		
		We assume a cooperation rate $R_2=0$, which means that Detector can send a message $M_2$ to Detector~2 that is described by  a sublinear number of bits. 
		
%Detector~1 sends a single bit to Detector 2, which is optimal as the following proposition shows.
The simple scheme in the next subsection \ref{sec:coding} achieves the following exponents region, which can be proved to be optimal.
		
		\begin{theorem}[Testing Against Independence]~\label{theorem-lower-bounds-power-exponents-test-against-independence-positive-rates}
			Assume $\bar h_1=1$ and \eqref{eq: condition case 2}. 
			Then, $\mathcal{E}(R_1,0)$ is the set of all nonnegative exponent pairs $(\theta_1,\theta_2)$  for which
			\begin{subequations}\label{power-exponent-test-vs-indep-with-indep-si-zero-conf}
				\begin{align}
				\label{power-exponent-test-vs-indep-with-indep-si-Decoder1-zero-conf}
				\theta_1 & \leq  I\left(U;Y_1\right) \\
				\theta_2 & \leq  I\left(U;Y_1\right) + I\left(U;Y_2\right),
				\label{power-exponent-test-vs-indep-with-indep-si-Decoder2-zero-conf}
				\end{align} 
			\end{subequations}for some $U$ satisfying the Markov chain $U\mkv X \mkv (Y_1,Y_2)$ and the rate constraint $R_1 \geq I(U;X)$. % that $(U,\textnormal{constant}) \in \mc S(R_1,R_2=0)$.
		\end{theorem} 
	
		\begin{IEEEproof}The achievability follows by specializing and evaluating Theorem~\ref{theorem-lower-bounds-power-exponents-general-hypotheses-positive-rates} for this setup. The converse is proved  in~Appendix~\ref{App-E}.
		\end{IEEEproof}
	\begin{lemma}[Cardinality bound]  the right hand sides of \eqref{power-exponent-test-vs-indep-with-indep-si-zero-conf} in Theorem~ \ref{theorem-lower-bounds-power-exponents-test-against-independence-positive-rates} remain valide if we impose the cardinality bound $\|\mc U\| = \|\mc X\| +1$.
	\end{lemma} 
\begin{IEEEproof}see \cite[Theorem~3]{H87}.
	\end{IEEEproof}
	
	Notice that for $R_2=0$ the scheme in the following subsection sends only a single bit and that without cooperation, the term $I\left(U;Y_2\right)$ needs to be removed on the right-hand side of \eqref{power-exponent-test-vs-indep-with-indep-si-Decoder2-zero-conf}. This mutual information term thus represents the benefit of a single cooperation bit from Detector 1 to Detector 2.
		
		We illustrate the benefit of cooperation  at hand of the following example. 
		\begin{example} \label{ex1}Consider a setup with coherent detection, $\bar{h}_1=1$, where $X, Y_1, Y_2$ are ternary  and under $\mc H=0$:
			\begin{align}
			\left\{\begin{array}{cccc} P_{X Y_1Y_2}(0,0,0) = 0.0250   & P_{X Y_1Y_2}(0,0,1) =0.0250 & P_{X  Y_1Y_2}(0,1,0) = 0.15& P_{X  Y_1Y_2}(0,1,1) =0.2250 \\
			P_{X  Y_1Y_2}(1,0,0) = 0.0250   & P_{X  Y_1Y_2}(1,0,1) =0.2000 & P_{X   Y_1Y_2}(1,1,0) = 0.0500& P_{X  Y_1Y_2}(1,1,1) =0.0125 \\
			P_{X  Y_1Y_2}(2,0,0) = 0.2000   & P_{X  Y_1 Y_2}(2,0,1) =0.0250 & P_{X  Y_1Y_2}(2,1,0) = 0.0500& P_{X  Y_1Y_2}(2,1,1) =0.0125  \end{array} \right.
			\end{align}
			whereas under $\mc H = 1$ they are independent with same marginals as under $\mc H=0$.
			\begin{figure}[!ht]
				\centering
				\includegraphics[width=10cm]{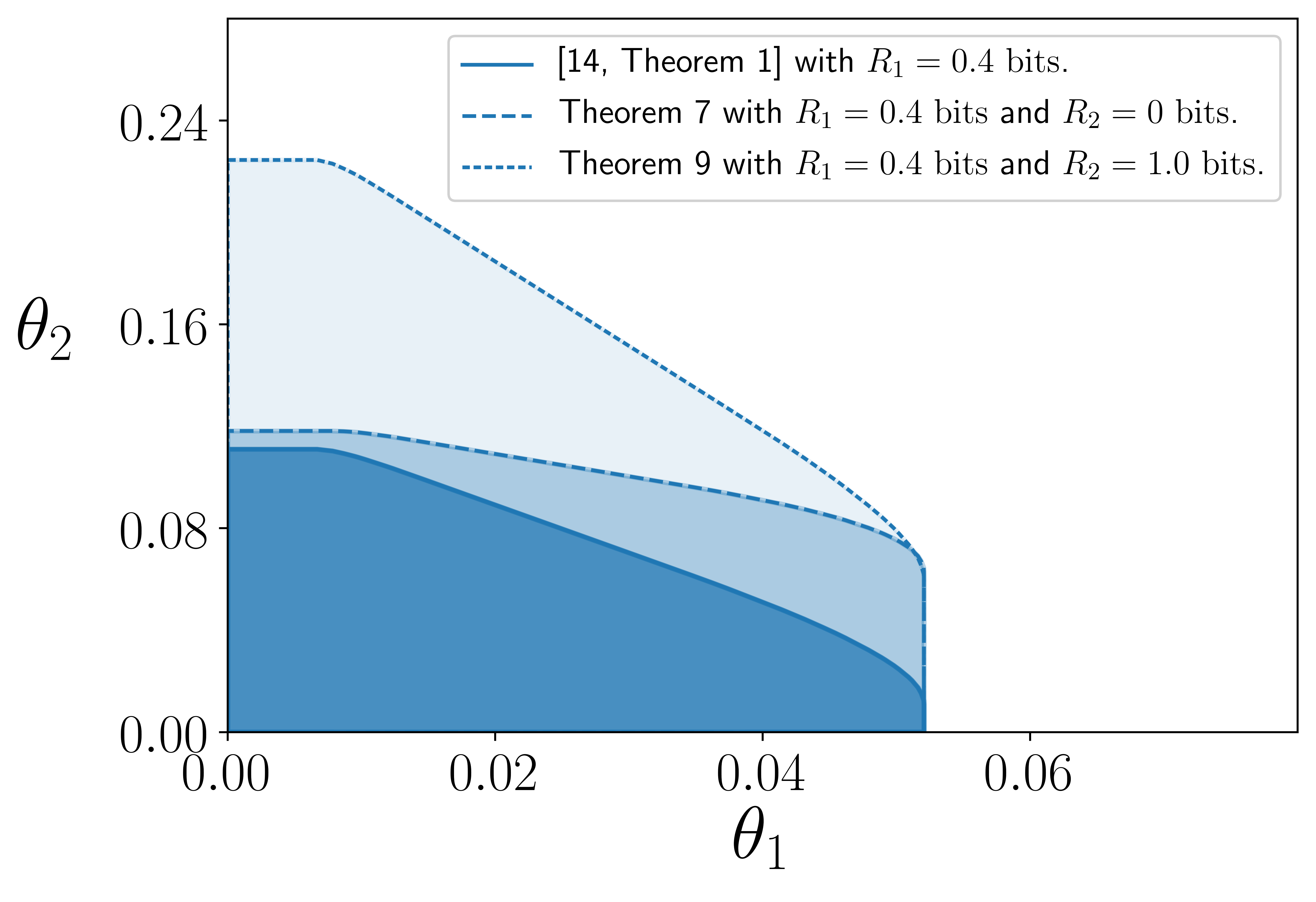}\vspace{-0.5cm}
				\caption{Error-exponents region of Example~\ref{ex1}, see \cite{EZW19C} for implementation details.}
				\label{fig:exponent-region-2}
			\end{figure}
			Fig.~\ref{fig:exponent-region-2} illustrates an achievable error-exponents region obtained with theorem~\ref{theorem-lower-bounds-power-exponents-general-hypotheses-positive-rates} when the communication rate are $R_1=0.4\textnormal{ bits}$ and $R_2=1.0 \textnormal{ bits}$. It also shows the error-exponents region $\mc E(0.4,0)$ presented in Theorem~\ref{theorem-lower-bounds-power-exponents-test-against-independence-positive-rates} and the error-exponents region without cooperation when $R_1=0.4\textnormal{ bits}$ derived in \cite[Theorem~1]{WT16}. Notice that the scheme achieving  $\mc E(0.4,0)$ requires that only a single cooperation bit is sent from Detector  1 to Detector 2.
		%	{\color{red}Please add an achievable region for $R_2 >0$. And increase the font on the axis of the figure.}
		\end{example}

\subsection{A Simple Scheme with Cooperation}\label{sec:coding}
In this subsection, we present coding schemes for both coherent and concurrent detection. Our schemes do not use binning or Heegard-Berger coding because of the complexity of the expressions describing the error exponents achieved with such schemes. In fact, such an approach would lead to expressions with 18 competing exponents,  {see the supplementary material for the case without cooperation.} % suggest to put the writeup for the binning scheme on a webpage and cite it here.}. 
Notice that this difficulty seems inherent to all multi-user hypothesis testing scenarios, see e.g. \cite{SWW17}. % {\color{red}do we have more examples?} {\color{blue} Abdellatif told me about one article from Han in Source coding I think with 11 exponent. I don't think it's only in HT}
We first present a  scheme  for coherent detection, 
\begin{equation}
\bar{h}_1=1,
\end{equation} and  then explain how to change the scheme for concurrent detection.

\subsubsection{Coherent Detection}

\underline{\textit{Preliminaries:}}
%The code construction differs on whether $P_X = \bar{P}_X$ or not and on the value of $h$ (i.e., whether we have coherent or concurrent detetion). 
%If 
%\begin{equation}
%P_X \neq  \bar{P}_X \qquad \textnormal{and } \qquad h=1  
%\end{equation}
Fix a  small $\delta>0$ and a pair of auxiliary random variables $(U, V)$ so that the following Markov chains 
\begin{IEEEeqnarray}{rCl}
	U \mkv X \mkv (Y_1, Y_2 )
%	\\
%	U_2 \mkv X \mkv (Y_1, Y_2 )
	\\ V\mkv (Y_1,U)\mkv (Y_2,X)
\end{IEEEeqnarray}
%When  $P_X \neq  \bar{P}_X$, then $\mu>0$ should be chosen so that the intersection 
%\begin{equation}\label{eq:inter}
%\mathcal{T}_{\mu/8}^n(P_{X}) \cap \mathcal{T}_{\mu/8}^n(\bar{P}_{X}) = \emptyset
%\end{equation}
%is empty. Fix also  a triple of auxiliary random variables $(U_1,U_2, V)$ so that the following Markov chains 
%\begin{IEEEeqnarray}{rCl}
%		U_1 \mkv X \mkv (Y_1, Y_2 )
%		\\
%	U_2 \mkv X \mkv (Y_1, Y_2 )
%	\\ V \mkv (Y_1,U_{h_2})\mkv (Y_2,X),
%\end{IEEEeqnarray}
and satisfying the rate constraints
\begin{align}
R_1 &>  I(U;X) \label{eq:rate1}\\
%R_1 &\geq  I(U_2;X) + \xi(\mu) \\
R_2 &>  I(V;Y_1|U).
\end{align}
%are satisfied
%where $\xi(\cdot ) \to 0$ is a function that tends to 0 as its argument tends  to 0. 

\underline{\textit{Codebook Generation:}} we randomly generate the codebook  
\begin{equation}
\mc C_{U} \triangleq\big\{ u^n(m_1) \colon  m_1 \in \{1,\ldots,\lfloor 2^{nR_1} \rfloor \}\big\}
\end{equation} by drawing each entry of each codeword $u^n(m_1)$ i.i.d. according to  $P_{U}$. 

Furthermore, we superpose a codebook $\mc C_{V}$ on codebook $\mc C_{U}$. %Notice that there is no need to superposition any codebook on $\mc C_{U,\bar{h}_2}$. % because when the Sensor decides to use codebook $C_{U,\bar{h}_2}$, Detector 2 immediately decides on its alternative hypothesis $\bar{h}_2$ without further processing the message $M_2$ obtained from Detector 1. 
So, for each index $m_1 \in \{1,\ldots,\lfloor 2^{nR_1} \rfloor\}$, we randomly  construct the codebook 
 \begin{equation} \mc C_{V} (m_1)\triangleq \{v^n(m_2|m_1) \colon m_2 \in \{1,\ldots,\lfloor 2^{nR_2} \rfloor\}\}\end{equation} by drawing the $j$-th entry of each codeword $v^n(m_2|m_1)$ according to the conditional pmf $P_{V|U}(\cdot|u_{j}(m_1))$, where  $u_{j}(m_1)$ denotes the $j$-th component of codeword $u^n(m_1)$.

Reveal all  codebooks to all terminals.

%As we will see, when
%\begin{equation}\label{eq:coherent}
%P_X =  \bar{P}_X \qquad \textnormal{or} \qquad \bar{h}_2=1, 
%\end{equation}
%then the random variable $U_{\bar{h}_2}$ and the codebook $\mathcal{C}_{U,\bar{h}_2}$  are never used and become meaningless. 
%

\underline{\textit{Sensor:}} Assume  it observes the source sequence $X^n=x^n$. Then, it first looks for a message $m_1\in \{1,\ldots,\lfloor 2^{nR_1} \rfloor\}$ such that 
\begin{equation}
(u^n(m_1), x^n) \in \mc T^n_{\mu/8}(P_{UX}).
\end{equation}If one or multiple such indices $m$  are found, the Sensor selects $m^*$ uniformly at random over these indices and sends 
\[ M_1 = (0, m_1^*).
\]
%over the noise-free link to both Detectors.
%(Note that the measure $P_{UX}$ is the one induced by the null hypothesis $\mc H$). 
%Otherwise, if 
%\begin{equation}P_X \neq \bar{P}_X \qquad \textnormal{and} \qquad \bar{h}_2=1,
%\end{equation} it looks for a message   $m\in \{1,\ldots,\lfloor 2^{nR_1} \rfloor\}$ such that 
%\begin{equation}\label{eq:check1}
%(u_{0}^n(m), x^n) \in \mc T^n_{\mu/8}(P_{U_0X}).
%\end{equation}f one or multiple such indices $m$  are found, the Sensor selects $m^*$ uniformly at random over these indices and sends 
%\[ M_1 = (0, m^*).
%\]
Otherwise, it sends 
\begin{equation}
M_1=0.
\end{equation}

%If one no such pair $(i,m)$ is found, the Sensor sends the index $M_1=0$ over the common noise-free pipe to both Detectors. Otherwise, it selects $(i^*,m^*)$ uniformly at random over all the pairs satisfying \eqref{eq:check1}  and sends $M_1=(i^*,m^*)$ to the two Detectors. (Notice that by \eqref{eq:inter}, when $P_X \neq \bar{P}_X$, there is only one value of $i$ for which \eqref{eq:check1} can be satisfied.)

\underline{\textit{Detector 1:}} If 
\begin{equation}M_1= 0%\qquad \textnormal{or} \qquad M_1=(1,m) \;\;\; \textnormal{for some }m\in \{1,\ldots,\lfloor 2^{nR_1} \rfloor\},
\end{equation} Detector 1 decides on the  alternative hypothesis 
\begin{equation}\label{eq:alt1}
\hat{\mc H}_1=1.
\end{equation} If 
\begin{equation}
M_1=(1,m_1)\quad \textnormal{for some }m_1 \in \{1,\ldots,\lfloor 2^{nR_1} \rfloor\},
\end{equation}
and given that  $Y_1^n= y_1^n$, Detector 1 checks whether 
\begin{equation}\label{eq:test1}
(u^n(m_1), y_1^n) \in \mc T^n_{\mu/4}(P_{UY}).
\end{equation} If the test is successful, it decides on the null hypothesis
\begin{equation}\label{eq:hyp}
\hat{\mc H}_1= 0.
\end{equation}
Otherwise it decides on the alternative hypothesis as in \eqref{eq:alt1}. %$\hat{\mc H}_1=\bar{h}_1$.

We now describe the communication to Detector 2. If \begin{equation}\label{eq:dd12}
\hat{\mc H}_1= 1
\end{equation} Detector 1 sends 
\begin{equation}
M_2=0.
\end{equation}
Otherwise, it looks for an index $m_2\in\{1,\ldots,\lfloor 2^{nR_2} \rfloor\}$ such that 
\begin{equation}(u^n(m_1), v^n(m_2|m_1), y_1^n) \in \mc T^n_{\mu/2}(P_{UVY_1}).\label{eq:test}
\end{equation}
If one or more such indices can be  found, Detector 1 selects an index $m_2^*$ among them uniformly at random and sends 
\begin{equation}M_2=m_2^*.\end{equation} Otherwise it sends $M_2=0$.

\underline{\textit{Detector 2:}} If \begin{equation}
M_1=0\qquad \textnormal{or} %\qquad M_1=(\bar{h}_2,m) \;\;\; \textnormal{for some }m\in \{1,\ldots,\lfloor 2^{nR_1} \rfloor\}\qquad \textnormal{or}
 \qquad M_2=0,
\end{equation}  Detector 2 decides on the alternative  hypothesis 
\begin{equation}\label{eq:alt2}\hat{\mc H}_2=1.
\end{equation} If
\begin{equation}M_1=(1,m_1) \;\;\; \textnormal{for some }m_1\in \{1,\ldots,\lfloor 2^{nR_1} \rfloor\} \textnormal{ and }  M_2=m_2 \;\;\; \textnormal{for some }m_2\in \{1,\ldots,\lfloor 2^{nR_2} \rfloor\},\end{equation}  and given  $Y_2^n=y_2^n$,  Detector 2 checks whether 
\begin{equation}
(u^n(m_1), v^n(m_2|m_1), y_2^n) \in \mc T^n_{\mu}(P_{UVY_2}).\label{eq:test2}
\end{equation} If this check is successful, Decoder~2 decides on the null hypothesis
\begin{equation}\label{eq:H20}
\hat{\mc H}_2=0.
\end{equation} Otherwise, it decides on the alternative hypothesis
\begin{equation}\hat{\mc H}_2=1.
\end{equation}

\subsubsection{Changes for concurrent detection when \texorpdfstring{$P_X =\bar{P}_X$}{PX = PXB} }
We now consider the scenario of concurrent detection, so 
\begin{equation}
\bar{h}_1=0.
\end{equation}
We  apply the  same scheme as above, except for the decision at Detector 1, which is described next.

\underline{\textit{Detector 1:}} 
 If 
 \begin{equation}M_1=0 %\qquad \textnormal{or} \qquad M_1=(1,m) \;\;\; \textnormal{for some }m\in \{1,\ldots,\lfloor 2^{nR_1} \rfloor\},
 \end{equation} Detector 1 now decides 
 \begin{equation}\label{eq:alt1alt}
 \hat{\mc H}_1=0.
 \end{equation} If 
 \begin{equation}
 M_1=(1,m_1)\quad \textnormal{for some }m_1\in \{1,\ldots,\lfloor 2^{nR_1} \rfloor\},
 \end{equation}
 and given that  $Y_1^n= y_1^n$, Detector 1 checks whether 
 \begin{equation}
 (u^n(m_1), y_1^n) \in \mc T^n_{\mu/4}(\bar{P}_{UY_1}).
 \end{equation} If the test is successful, it decides 
 \begin{equation}\label{eq:hyp-b}
 \hat{\mc H}_1= 1.
 \end{equation}
 Otherwise it decides 
 \begin{equation}
 \hat{\mc H}_1=0.
 \end{equation}

Communication to Detector 2 is as described in the previous subsection.

\subsubsection{Changes for concurrent detection when \texorpdfstring{$P_X \neq\bar{P}_X$}{PX not PXB}}

If $\bar{h}_1=0$ and   $P_X \neq\bar{P}_X$, the scheme should be changed as described in the previous paragraph. The following additional changes allow to obtain an improved scheme.

In this case, we choose $\mu>0$  so that the intersection 
\begin{equation}\label{eq:inter}
\mathcal{T}_{\mu/8}^n(P_{X}) \cap \mathcal{T}_{\mu/8}^n(\bar{P}_{X}) = \emptyset
\end{equation}
is empty and we choose another  auxiliary random variable $\bar{U}_1$ satisfying 
\begin{IEEEeqnarray}{rCl}
%		U_1 \mkv X \mkv (Y_1, Y_2 )
%		\\
	\bar{U}_1 \mkv \bar{X} \mkv (\bar{Y}_1, \bar{Y}_2 )\\
%	\\ V \mkv (Y_1,U_{h_2})\mkv (Y_2,X),
%R_1 &\geq  I(U_0;X) + \xi(\mu) \\
R_1 \geq  I(\bar{U}_1;\bar{X}) + \xi(\mu).
%R_2 &\geq  I(V_0;Y_1|U_{0}) + \xi(\mu).
\end{IEEEeqnarray}
%are satisfied
%where $\xi(\cdot ) \to 0$ is a function that tends to 0 as its argument tends  to 0. 
A third codebook 
\begin{IEEEeqnarray}{rCl}
\mc C_{U,1} \triangleq\big\{ u_1^n(m_1) \colon  m \in \{1,\ldots,\lfloor 2^{nR_1} \rfloor \}\big\}
\end{IEEEeqnarray} is  drawn by picking the entries i.i.d. according to  $\bar{P}_{U_1}$.

Encoding has to be changed as follows. If the test in \eqref{eq:test} fails, then the Sensor looks for an index  $m_1\in \{1,\ldots,\lfloor 2^{nR_1} \rfloor\}$ such that 
\begin{equation}\label{eq:check1}
(u_{1}^n(m_1), x^n) \in \mc T^n_{\mu/8}(\bar{P}_{U_{1}X}).
\end{equation}
If one or multiple such indices $m_1$  are found, the Sensor selects $m_1^*$ uniformly at random over these indices and sends 
\begin{equation}\label{eq:M1} M_1 = (2 , m_1^*)
\end{equation}
Notice that by the condition \eqref{eq:inter}, only one of the two tests \eqref{eq:test} and \eqref{eq:check1} can be successful for any observed sequence $x^n$. It therefore does not matter which one is performed first.

Thus, now the Sensor sends three different types of messages: 
\begin{equation}
M_1 =0 \qquad \textnormal{or} \qquad M_1 =(1,m_1) \qquad \textnormal{or} \qquad M_1 =(2,m_1).
\end{equation}
The message $M_1=(1,m_1)$ indicates  that  the Sensor is tempted to guess $\mathcal{H}=0$. After receiving such a message, Detector 1 therefore  produces $\hat{\mathcal{H}}_1=0$. The same holds if $M_1=0$. In contrast, if $M_1=(2,m_1)$, Detector 1 checks whether 
\begin{equation}\label{eq:test1-b}
(u_{1}^n(m_1), y_1^n) \in \mc T^n_{\mu/4}(\bar{P}_{U_{1}Y_1}).
\end{equation}  
If successful it declares $\hat{\mathcal{H}}_1=1$, and otherwise $\hat{\mathcal{H}}_1=0$. 

Communication from, Detector 1 to Detector 2 is as described before.

Similarly, the message $M_1=(2,m_1)$ now indicates  that  the Sensor is tempted to guess $\mathcal{H}=1$. When receiving this message,  Detector 2 therefore decides immediately  $\hat{\mathcal{H}}_2=1$.  Otherwise it  acts as described in the original scheme. 
\subsection{Achievable regions}
We now present the  regions  achieved by the coding scheme described in the previous subsection. Notice that the new achievable regions recover the extreme cases in the previous section \ref{sec:large}, when the rates are set accordingly. 

	We first consider coherent detections $\bar h_1 = 1$.	
			
		For given rates $R_1 \geq 0$ and $R_2 \geq 0$, define the following set of auxiliary random variables:
		\begin{equation} \mathcal{S} \left(R_1,R_2\right) \triangleq \left\{ \left(U,V\right) \colon \left. \begin{array}{c} U \mkv X \mkv (Y_1, Y_2 )\\ V \mkv( Y_1,U )\mkv (Y_2,X) \\ I\left(U;X\right) \leq R_1 \\ I\left(V;Y_1|U \right) \leq R_2 \end{array} \right.\right\}.\end{equation}
			Further, define for each $(U,V) \in \mathcal{S} \left(R_1,R_2\right)$, the sets 
			\begin{IEEEeqnarray}{rcl}	
						\mathcal{L}_1\left(U\right)& \triangleq &\left\{(\tilde{U},\tilde{X},\tilde{Y}_1) \colon \begin{array}{c}   P_{\tilde{U}\tilde{X}} = P_{UX}\\ P_{\tilde{U}\tilde{Y}_1} = P_{UY_1} \end{array}\right\}\label{eq:L1}
%						\mathcal{L}_{\text{nc},2}\left(U\right) &\triangleq &\left\{(\tilde{U},\tilde{X},\tilde{Y}_2) \colon  \begin{array}{c}   P_{\tilde{U}\tilde{X}} = P_{UX}\\ P_{\tilde{U}\tilde{Y}_2} = P_{UY_2} \end{array} \right\}. 
					\end{IEEEeqnarray}
%		\begin{align*}
%		\mathcal{L}_1\left(U\right) \triangleq \left\{(\tilde{U},\tilde{X},\tilde{Y}_1) \colon P_{\tilde{U}\tilde{X}} = P_{UX},\  P_{\tilde{U}\tilde{Y}_1} = P_{UY_1}\right\},
%		\end{align*}
		and
		\begin{align}
		\mathcal{L}_2\left(UV\right) &\triangleq \left\{(\tilde{U},\tilde{V},\tilde{X},\tilde{Y}_1,\tilde{Y}_2) \colon  \begin{array}{c}   P_{\tilde{U}\tilde{X}} = P_{UX}\\  P_{\tilde{U}\tilde{V}\tilde{Y}_1} = P_{UVY_1}\\  P_{\tilde{U}\tilde{V}\tilde{Y}_2} = P_{UVY_2} \end{array} \right\},
		\end{align}
	%	Also, let  $(\bar{X}, \bar{Y}_1, \bar{Y}_2) \sim \bar P_{X Y_1Y_2}$ and 
		and  the random variables $(\bar{U}, \bar{V})$ so as to satisfy 
		\begin{equation}
P_{\bar{U}|\bar{X}}=P_{U|X} \quad \textnormal{and} \quad P_{\bar{V}|\bar{Y_1}\bar{U}}=P_{V|Y_1U}
		\end{equation}
		 and the Markov chains
		\begin{IEEEeqnarray}{rCl}
		\bar{U} \mkv \bar{X} \mkv (\bar{Y}_1,\bar{Y}_2) \\
		\bar{V} \mkv (\bar{Y_1},\bar{U}) \mkv (\bar{X},\bar{Y}_2).
		\end{IEEEeqnarray}
	
		\begin{theorem}[Coherent Detection]~\label{theorem-lower-bounds-power-exponents-general-hypotheses-positive-rates} 
If 
					\begin{equation} 	 \bar{h}_1 = 1,\end{equation} 
the exponents region $\mathcal{E}(R_1, R_2)$ contains all nonnegative  pairs $(\theta_1,\theta_2)$ that satisfy:
			\begin{IEEEeqnarray}{rcl}\subnumberinglabel{eq: an achievable rate exponent region for Heegard-Berger}
			\theta_1 & \leq & \min_{\tilde{U}\tilde{X}\tilde{Y}_1 \in \mathcal{L}_1\left(U\right)} D\left(\tilde{U}\tilde{X}\tilde{Y}_1||\bar{U}\bar{X}\bar{Y}_1\right)  \label{eq: an achievable rate exponent region for Heegard-Berger Detector 1} \\
			\theta_2&  \leq &\min_{\tilde{U}\tilde{V}\tilde{X}\tilde{Y}_1\tilde{Y}_2 \in \mathcal{L}_2\left(UV\right)} D\left(\tilde{V}\tilde{U}\tilde{X}\tilde{Y}_1\tilde{Y}_2||\bar{V}\bar{U}\bar{X}\bar{Y}_1\bar{Y}_2\right).\label{eq: an achievable rate exponent region for Heegard-Berger Detector 2}
			\end{IEEEeqnarray}
			for some $(U,V) \in \mathcal{S} \left(R_1,R_2\right)$ 
		\end{theorem}
		\begin{IEEEproof}
			The exponent region is achieved by the scheme described in Subsection~\ref{sec:coding}. The proof is given in Appendix~\ref{secV_subsecA}. 
		\end{IEEEproof}
		
	For our second result, we also define for each auxiliary random variable $U$ the set
			\begin{IEEEeqnarray}{rcl}	
				\overline{\mathcal{L}}_1\left(U\right)& \triangleq &\left\{(\tilde{U},\tilde{X},\tilde{Y}_1) \colon \begin{array}{c}   P_{\tilde{U}\tilde{X}} = \bar P_{UX}\\  P_{\tilde{U}\tilde{Y}_1} = \bar P_{ UY_1} \end{array}\right\}.\label{eq:L1bar}
			\end{IEEEeqnarray}

					\begin{theorem}[Concurrent Detection with $P_X =\bar{P}_X$]\label{thm:concurrent1}
						If 
						\begin{equation} 	 \bar{h}_1 = 0, \text{ and } P_X=\bar P_X ,
						\end{equation}then the exponents region $\mathcal{E}(R_1, R_2)$ contains all nonnegative  pairs $(\theta_1,\theta_2)$ that for some $(U,V) \in \mathcal{S} \left(R_1,R_2\right)$ satisfy:
						\begin{IEEEeqnarray}{rcl}\subnumberinglabel{eq: an achievable rate exponent region for Heegard-Berger-concurrent}
							\theta_1 & \leq & \min_{\tilde{U}\tilde{X}\tilde{Y}_1 \in \overline{\mathcal{L}}_1\left(U\right)} D\left(\tilde{U}\tilde{X}\tilde{Y}_1||{U}{X}{Y}_1\right)  \label{eq: an achievable rate exponent region for Heegard-Berger-concurrent Detector 1} \\
							\theta_2&  \leq &\min_{\tilde{U}\tilde{V}\tilde{X}\tilde{Y}_1\tilde{Y}_2 \in \mathcal{L}_2\left(UV\right)} D\left(\tilde{V}\tilde{U}\tilde{X}\tilde{Y}_1\tilde{Y}_2||\bar{V}\bar{U}\bar{X}\bar{Y}_1\bar{Y}_2\right).\label{eq: an achievable rate exponent region for Heegard-Berger-concurrent Detector 2}
						\end{IEEEeqnarray}
					\end{theorem}
					\begin{IEEEproof}
						Similar to the proof of Theorem~\ref{theorem-lower-bounds-power-exponents-general-hypotheses-positive-rates} and omitted. 
					\end{IEEEproof}
					
		For a  given rate $R_1 \geq 0$, define the following set of auxiliary random variables:
		\begin{equation} \mathcal{S} \left(R_1\right) \triangleq \left\{ \bar  U_1 \colon \left. \begin{array}{c} \bar{U}_1 \mkv \bar{X} \mkv \bar{Y}_1\\ I\left(\bar{U}_1;\bar{X}\right) \leq R_1 \end{array} \right.\right\}.\end{equation}
		and the random variable $U_1$ so that $P_{U_1|X}=\bar{P}_{U_1|X}$ and the Markov chain $U_1 \mkv X \mkv Y_1$ holds. 
		
							\begin{theorem}[Concurrent Detection and $P_X \neq \bar{P}_X$] \label{thm:concurrent}
								If 
								\begin{equation} 	 \bar{h}_1 = 0 \qquad \textnormal{and} \qquad P_X \neq \bar{P}_X,\end{equation} 
								then the exponents region $\mathcal{E}(R_1, R_2)$ contains all nonnegative  pairs $(\theta_1,\theta_2)$ that for some $(U,V) \in \mathcal{S} \left(R_1,R_2\right)$ and $\bar{U}_1 \in \mathcal{S}_{\textnormal{nc}}(R_1)$ satisfy:
								\begin{IEEEeqnarray}{rcl}
									\theta_1 & \leq & \min_{\tilde{U}_1\tilde{X}\tilde{Y}_1 \in \overline{\mathcal{L}}_1\left(\bar U_1\right)} D\left(\tilde{U}_1\tilde{X}\tilde{Y}_1||{U}_1{X}{Y}_1\right)  \label{eq: an achievable rate exponent region for Heegard-Berger-concurrent-diff-in-marginal Detector 1} \\
									\theta_2&  \leq &\min_{\tilde{U}\tilde{V}\tilde{X}\tilde{Y}_1\tilde{Y}_2 \in \mathcal{L}_2\left(UV\right)} D\left(\tilde{V}\tilde{U}\tilde{X}\tilde{Y}_1\tilde{Y}_2||\bar{V}\bar{U}\bar{X}\bar{Y}_1\bar{Y}_2\right).\label{eq: an achievable rate exponent region for Heegard-Berger-concurrent-diff-in-marginal Detector 2}
								\end{IEEEeqnarray}
							\end{theorem}
							\begin{IEEEproof}
			The proof is given in Appendix~\ref{app:concurrent}. It is based on the scheme of the previous Subsection~\ref{sec:coding}.
							\end{IEEEproof}
				
\begin{remark}
As seen in Theorem~\ref{theorem-lower-bounds-power-exponents-test-against-independence-positive-rates}, in some special case exponents accumulates. 
\end{remark}
		\begin{remark}
			The exponents region in Theorem~\ref{thm:concurrent} is rectangular because $\theta_1$ depends only on the auxiliary $\bar{U}_1$ and $\theta_2$ only on the pair of auxiliaries $(U,V)$. This implies that both exponents can be maximized at the same time without any tradeoff between the two exponents. 
			
			This  is  different in the first two Theorem~\ref{theorem-lower-bounds-power-exponents-general-hypotheses-positive-rates} and \ref{thm:concurrent1} where both exponents depend on the same auxiliary, and therefore the regions exhibit a tension when maximizing the two exponents.
		\end{remark}

\section{Summary and Conclusion}\label{sec:conclusion}
In this paper we  investigated the role of cooperation under both coherent and concurrent detection in a two-detector hypothesis testing system. We characterized fully the exponents region for fixed communication alphabets and in a testing-against independence scenario with positive  communication rates. For the general positive-rate scenario, we proposed a simple scheme in which Detector 1 uses the cooperation link to inform Detector 2 about its guess and a compressed version of its observations. Our scheme behaves differently for coherent and for concurrent detection. In fact, for coherent detection, if possible the sensor guesses the hypothesis, conveys this guess to the detectors, and then focuses on helping the detector that wishes to maximize the exponent under the  hypothesis not corresponding to its guess.

%We have then derived lower bound on the exponent of type II error for cases of general hypothesis and gives optimality results in a case of testing against independence. 
Our results allowed us to exactly quantify the gains of cooperation for communication with finite alphabets  and for the  testing against independence scenario solved in this paper. Cooperation gains are not bounded even with only a single bit of cooperation.

Depending on the specific scenario (coherent or concurrent detection, positive or zero rate,  $P_X = \bar{P}_X$ or $P_X \neq \bar{P}_X$), the exponents achieved at the two detectors may or may not show a tradeoff. The absence of such a tradeoff implies that each detector can achieve the same exponent as if it were the only detector in the system. %Notice that all these results generalize for any finite number of hypothesis.
\section*{Acknowledgement}
M. Wigger acknowledges funding support from the European Research Council under grant agreement 715111.
\appendices% \label{secV}
\section{Proof of the Converse Bound \eqref{eq:converse2}  in Proposition~\ref{prop-coherent}}~\label{app:converse2}

Fix an achievable Type-II error exponent $\theta_2$. Then choose  a small number $\epsilon>0$, a sufficiently large blocklength $n$,  and encoding and decision functions $\phi_{1,n}, \phi_{2,n}, \psi_{2,n}$ satisfying 
\begin{equation}\label{eq-type-I-constraint-D2-p}
\alpha_{2,n} \leq \epsilon
\end{equation} 
and 
\begin{equation}
-\frac{1}{n} \log \beta_{2,n} \geq \theta_2- \epsilon. 
\end{equation}
For the chosen encoding and decision functions, define  for each pair $(m_1,m_2) \in \{0,\ldots,\W_1-1\} \times \{ 0,\ldots,\W_2-1\}$ the subsets 
\begin{IEEEeqnarray}{rCl}
	\mc C_{m_1}& \triangleq &\{ x^n \in \mc X^n \colon  \phi_{1,n}(x^n) =m_1\} ,\\
	\mc G_{m_1,m_2}& \triangleq &\{ y_1^n \in \mc Y_1^n \colon    \phi_{2,n}(m_1,y_1^n)=m_2 \} ,\\
		\mc F_{m_1,m_2}& \triangleq &\{ y_2^n \in \mc Y_2^n \colon   \psi_{2,n}(m_1,m_2,y_2^n) =0\} .
	\end{IEEEeqnarray}
	Notice that the sets
	$\mc {C}_{0}, \ldots, \mc C_{W_1-1}$ partition $\mathcal{X}^n$ and  for each $m_1 \in \{0,\ldots,\W_1-1\}$ the sets $\mc G_{m_1,0}, \ldots, \mc G_{m_1,\W_2-1}$ partition $\mathcal{Y}_1^n$.
Moreover, the  acceptance region $\mc A^2_n$ at Detector $2$, defined through the relation
\begin{equation}
 (X^n,Y_1^n,Y_2^n) \in \mc A^2_n  \Longleftrightarrow \hat {\mc H}_2 =0 ,
\end{equation} can  be expressed as
\begin{equation}\label{eq:type-I-constraint concurent-cooperation-0-p}
\mc A^2_n \triangleq \bigcup_{m_2=0}^\mathsf{W_2-1}\bigcup_{m_1 = 0}^{\W_1 -1} \mc C_{m_1} \times \mc G_{m_1,m_2} \times \mc F_{m_1,m_2}.
\end{equation}
By the constraint on the type-I error probability on  Detector $2$,  \eqref{eq-type-I-constraint-D2-p},
\begin{IEEEeqnarray}{rCl}
{ P_{XY_1Y_2}^{\otimes n}}\Bigg[\big ( X^n, Y^n_1,  Y^n_2\big ) \in \bigcup_{m_1=0}^{\W_1-1}\bigcup_{m_2=0}^{\W_2-1}{\mc C_{m_1}\times \mc G_{m_1,m_2} \times \mc F_{m_1,m_2}}   \Bigg] \geq 1 -\epsilon
\end{IEEEeqnarray}

Now, by the union bound there exists  an index pair $(m_1^*,m_2^*) \in \{0,\ldots,\W_1-1\} \times \{ 0,\ldots,\W_2-1\}$ such that:
\begin{subequations}\label{eq-type-I-constraint-concurent-cooperation-2-p}
	\begin{IEEEeqnarray}{rcl}
	 P_X^{\otimes n}\big[  X^n\in \mc C_{m_1^*} \big] &\ \geq\ & \frac{1- \epsilon}{\W_1}, \\
	 P^{\otimes n}_{Y_1}\big[  Y^n_1 \in \mc G_{m_1^*,m_2^*}\big] &\ \geq\ & \frac{1- \epsilon}{\W_1\W_2}, \\
	 P^{\otimes n}_{Y_2}\big[  Y^n_2 \in \mc F_{m_1^*,m_2^*}\big] &\ \geq\ & \frac{1- \epsilon}{\W_1\W_2},
	\end{IEEEeqnarray}
\end{subequations}

Combining \eqref{eq-type-I-constraint-concurent-cooperation-2-p} with an extension of \cite[Theorem 3]{SP92} to three pmfs (recall that we assumed   $P_{XY_1Y_2}(x,y_1,y_2)>0$ and thus $P_{XY_1Y_2} \ll \bar P_{XY_1Y_2}$), for sufficiently large $n$, we obtain:
\begin{align*}
\text{Pr}[ \hat{\mc H}_2=0 | \mc H =1  ]\geq  \max_{ \substack{\tilde P_{XY_1Y_2}: \\ \tilde P_{X} = P_X ,\\\tilde P_{Y_1} =  P_{Y_1}  ,  \tilde P_{Y_2}=P_{Y_2}}} e^{-n \big(D\big( \tilde P_{XY_1Y_2} \| \bar P_{XY_1Y_2}\big) + \mu\big)}. 
\end{align*}

 Taking $n \to \infty$ and $\mu \to 0$, by the continuity of KL-divergence, we can conclude  that for any achievable exponent $\theta_2$: 
	\begin{IEEEeqnarray}{rCl}
	\theta_{2} &\leq& \min_{ \substack{\tilde P_{XY_1Y_2}: \\ \tilde P_{X} = P_X ,\\ \tilde P_{Y_1} =  P_{Y_1} ,  \tilde P_{Y_2}=P_{Y_2}}}  D\Big( \tilde P_{XY_1Y_2} \| \bar P_{XY_1Y_2} \Big) \label{eq-bound-cooperative-2-p}. \IEEEeqnarraynumspace 
	\vspace{0.3cm}
	\end{IEEEeqnarray}
This conclude the proof.

\section{Converse To Theorem~\ref{theorem: optimal concurrent hypothesis testing zero rate with cooperation and depletion at the encoder}}\label{App-B}

Fix  a real number $r$ and an exponent pair $(\theta_1,\theta_2)\in \mathcal{E}_0(2,2)$ satisfying
\begin{equation}\label{eq:r}
\theta_2= \theta_1 +r.
\end{equation}
Then fix a small number $\epsilon>0$, a sufficiently large blocklength $n$,  and encoding and decision functions $\phi_{1,n}, \phi_{2,n}, \psi_{1,n},\psi_{2,n}$ satisfying 
 \begin{IEEEeqnarray}{rCl}
 	\alpha_{1,n} &\leq& \epsilon, \label{eq-type-I-constraint-D1}\\
 	\alpha_{2,n}& \leq &\epsilon, \label{eq-type-I-constraint-D2}
 \end{IEEEeqnarray} 
and
  \begin{IEEEeqnarray}{rCl}
-\frac{1}{n} \log \beta_{1,n} \geq \theta_1-\epsilon, \\
-\frac{1}{n} \log \beta_{2,n} \geq \theta_2-\epsilon.
 	\end{IEEEeqnarray}

 For the chosen encoding and decision functions, define for each $m_1 \in \{0,1,\ldots, \W_1-1\}$ and $m_2 \in \{0,1,\ldots, \W_2-1\}$, the subsets 
 \begin{IEEEeqnarray}{rCl}
 	\mc C_{m_1}& \triangleq &\{ x^n \in \mc X^n \colon  \phi_{1,n}(x^n) =m_1\},  \\
 	\mc F_{m_1}^1& \triangleq &\{ y_1^n \in \mc Y_1^n \colon    \psi_{1,n}(m_1,y_1^n)=1 \},\\	
 	\mc G_{m_1,m_2}& \triangleq &\{ y_1^n \in \mc Y_1^n \colon    \phi_{2,n}(m_1,y_1^n)=m_2 \} ,\\
 	\mc F_{m_1,m_2}^2& \triangleq &\{ y_2^n \in \mc Y_2^n \colon   \psi_{2,n}(m_1,m_2,y_2^n) =0\} 
 \end{IEEEeqnarray}
 Notice that the sets
 $\mc {C}_{0}, \ldots, \mc C_{W_1-1}$ partition $\mathcal{X}^n$ and  for each $m_1 \in \{0,\ldots,\W_1-1\}$ the sets $\mc G_{m_1,0}, \ldots, \mc G_{m_1,\W_2-1}$ partition $\mathcal{Y}_1^n$.
 Moreover, the  acceptance regions $\mc A^1_n$ and $\mc A^2_n$ at Detectors $1$ and $2$, defined through the relations
\begin{IEEEeqnarray}{C}
(X^n,Y_1^n) \in \mc A^1_n  \Longleftrightarrow \hat {\mc H}_1 =1 , \\(X^n,Y_1^n,Y_2^n) \in \mc A^2_n  \Longleftrightarrow \hat {\mc H}_2 =0  ,
\end{IEEEeqnarray} can be expressed as 
\begin{subequations}
	\label{eq:type-I-constraint concurent-cooperation-0}
	\begin{equation}
	\mc A^1_n = \mc C_0 \times \mc F_0^{1} \cup \mc C_1 \times \mc F_1^{1}
	\end{equation}
	and
	\begin{equation}
	\mc A^2_n = \bigcup_{m_2=0}^\mathsf{W_2-1}{\mc C_0\times \mc G_{0,m_2} \times \mc F_{0,m_2}^{2}} \cup \bigcup_{m_2=0}^\mathsf{W_2-1}{\mc C_1\times \mc G_{1,m_2} \times \mc F_{1,m_2}^{2}}.
	\end{equation}
\end{subequations}

Define now for each $m_1 \in \{0,1\}$ the set 
\begin{IEEEeqnarray}{rCl}\label{eq-phin-composite}
\Gamma_{m_1,n} := \bigg\{  \tilde{P}_X \in \mathcal{P}(\mathcal{X}) \colon \;  \tilde{P}_X^{\otimes n} \big[ X^n \in \mc C_{m_1}  \big] \geq \frac{1-\epsilon}{2}  \bigg\},\IEEEeqnarraynumspace
\end{IEEEeqnarray}
and for each pair $(m_1,m_2) \in \{0,1\} \times \{0,\ldots,\W_2-1\}$ the set 
\begin{IEEEeqnarray}{rCl}\label{eq-psin-composite}
	\Delta_{m_1,m_2,n} := \bigg\{  \tilde{P}_{Y_1} \in \mathcal{P}(\mathcal{Y}_1) \colon \;  \tilde{P}_{Y_1}^{\otimes n} \big[  Y^n_1 \in \mc G_{m_1,m_2}  \big] \geq \frac{1-\epsilon}{2\W_2}  \bigg\}.\IEEEeqnarraynumspace
\end{IEEEeqnarray}
Since the sets $\mathcal{C}_0, \mc C_1$ cover $\mathcal{X}^n$ and since for each $\tilde{P}_X\in\mathcal{P}(\mc X)$, it holds that $ \tilde{P}_X^{\otimes n} \big[ X^n \in \mc X^n \big] =1$, the  subsets $\Gamma_{0,n}, \Gamma_{1,n}$ cover the set $\mathcal{P}(\mathcal{X})$. Similarly, since for each $m_1 \in \{ 0,1\}$ the sets $\mc G_{m_1,0}, \ldots,  \mc G_{m_1,\W_2-1}$ cover $\mc Y_1^n$, the subsets $\Delta^{n}_{m_1,0}, \ldots, \Delta^{n}_{m_1,\W_2-1,n}$ cover the set $\mathcal{P}(\mathcal{Y}_1)$. 
Moreover, by the constraint on the type-I error probability at Detectors $1$ and $2$, \eqref{eq-type-I-constraint-D1} and \eqref{eq-type-I-constraint-D2}: 
\begin{IEEEeqnarray}{rCl}
	{ \bar P_{XY_1}^{\otimes n}}\Bigg[ \big ( X^n, Y^n_1\big ) \in \bigcup_{m_1=0}^\mathsf{1}{\mc C_{m_1} \times \mc F^1_{m_1}}   \Bigg] &\geq& 1 -\epsilon \\
	{ P_{XY_1Y_2}^{\otimes n}}\Bigg[\big ( X^n, Y^n_1,  Y^n_2\big ) \in \bigcup_{m_1=0}^{1}\bigcup_{m_2=0}^{\W_2-1}{\mc C_{m_1}\times \mc G_{m_1,m_2} \times \mc F_{m_1,m_2}^{2}}   \Bigg]& \geq &1 -\epsilon.
\end{IEEEeqnarray}

By the union bound there exist thus an index $\tilde m_1 \in \{0,1\}$ and an index pair $(m_1^*,m_2^*) \in \{0,1\} \times \{ 0,\ldots,\W_2-1\}$ such that:
\begin{subequations}\label{eq-type-I-constraint-concurent-cooperation-1}
	\begin{IEEEeqnarray}{rCl}
	\bar	{P}_X^{\otimes n}\big[ X^n\in \mc C_{\tilde m_1} \big] &\geq& \frac{1- \epsilon}{2}, \\
	\bar	P^{\otimes n}_{Y_1}\big[  Y^n_1 \in \mc  F_{\tilde m_1}^{1}\big] &\geq& \frac{1- \epsilon}{2},
	\end{IEEEeqnarray}
\end{subequations}
and
\begin{subequations}\label{eq-type-I-constraint-concurent-cooperation-2}
	\begin{IEEEeqnarray}{rcl}
		 P_X^{\otimes n}\big[  X^n\in \mc C_{m_1^*} \big] &\ \geq\ & \frac{1- \epsilon}{2}, \\
		 P^{\otimes n}_{Y_1}\big[  Y^n_1 \in \mc G_{m_1^*,m_2^*}\big] &\ \geq\ & \frac{1- \epsilon}{2\W_2}, \\
		 P^{\otimes n}_{Y_2}\big[  Y^n_2 \in \mc F^{2}_{m_1^*,m_2^*}\big] &\ \geq\ & \frac{1- \epsilon}{2\W_2},
	\end{IEEEeqnarray}
\end{subequations}
Combining \eqref{eq-type-I-constraint-concurent-cooperation-1} with the definition of $\Delta_{\tilde m_1,n}$ in \eqref{eq-phin-composite} and  with  \cite[Theorem 3]{SP92} (recall that by assumption $P_{XY_1}(x,y_1)>0$, for all $(x,y_1) \in \mathcal{X}\times \mathcal{Y}_1$) yields that for any $\mu>0$ and sufficiently large $n$ :
\begin{align*}
\text{Pr}[ \hat{\mc H}_1=1 | \mc H =0  ]\geq  \max_{ \substack{\tilde P_{XY_1}: \\ \tilde P_{X} \in \Gamma_{\tilde m_1,n} ,\\  \tilde P_{Y_1}=\bar P_{Y_1}}} e^{-n (D( \tilde P_{XY_1} \| P_{XY_1}) + \mu)}. 
\end{align*}

In the same way, combining \eqref{eq-type-I-constraint-concurent-cooperation-2} \eqref{eq-phin-composite} with \eqref{eq-psin-composite} and  extending \cite[Theorem 3]{SP92} to three pmfs (recall that by assumption $P_{XY_1Y_2}(x,y_1,y_2)>0$, for all $(x,y_1,y_2) \in \mathcal{X}\times \mathcal{Y}_1\times \mathcal{Y}_2$), for sufficiently large $n$:
\begin{align*}
\text{Pr}[ \hat{\mc H}_2=0 | \mc H =1  ]\geq  \max_{ \substack{\tilde P_{XY_1Y_2}: \\ \tilde P_{X} \in \Gamma_{m_1^*,n} ,\\\tilde P_{Y_1} \in \Delta_{m_1^*,m_2^*,n} ,\;  \tilde P_{Y_2}=P_{Y_2}}} e^{-n (D( \tilde P_{XY_1Y_2} \| \bar  P_{XY_1Y_2}) + \mu)}. 
\end{align*}
%Notice that by \eqref{eq-type-I-constraint-concurent-cooperation-1}, the p.m.f. $\bar P_X \in \Gamma__{\tilde m_1,n}$, by \eqref{eq-type-I-constraint-concurent-cooperation-2},  the p.m.f. $P_X \in \Delta_{m^*_1,n}$, and by \eqref{eq-type-I-constraint-concurent-cooperation-2}, the p.m.f. $P_{Y_1} \in \Delta_{m_1^*,m_2^*,n}$. 

Taking now $n \to \infty$ and $\mu \to 0$, by the continuity of the KL divergence we can conclude  that if the exponent pair $(\theta_1, \theta_2)$ is achievable, then there exist subsets $\Gamma_0,\Gamma_{1}$ that cover $\mc P(\mc X)$, subsets $\Delta_{0,0},\ldots, \Delta_{0,\W_2-1}$ that cover $\mc P(\mc Y_1)$, and subsets $\Delta_{1,0},\ldots, \Delta_{1,\W_2-1}$ that cover $\mc P(\mc Y_1)$  so that:
\begin{subequations}\label{eq-bound-cooperative}
	\begin{IEEEeqnarray}{rCl}
		\theta_{1} &\leq& \min_{ \substack{\tilde P_{XY_1}: \\ \tilde P_{X} \in \Gamma_{b} ,\\  \tilde P_{Y_1}= \bar P_{Y_1}}}  D\Big( \tilde P_{XY_1} \| P_{XY_1} \Big) , \IEEEeqnarraynumspace \label{eq-bound-cooperative-1} \\
		\vspace{0.3cm}
		\theta_{2} &\leq& \min_{ \substack{\tilde P_{XY_1Y_2}: \\ \tilde P_{X} \in \Gamma_{c} ,\\ \tilde P_{Y_1} \in \Delta_{ c,c_2} ,\;  \tilde P_{Y_2}=P_{Y_2}}}  D\Big( \tilde P_{XY_1Y_2} \| \bar P_{XY_1Y_2} \Big) \label{eq-bound-cooperative-2}. \IEEEeqnarraynumspace 
		\vspace{0.3cm}
	\end{IEEEeqnarray}
\end{subequations}
where the indices $b,c \in \{0,1\}$ and $c_2\in\{0,\ldots, \W_2-1\}$ are such that 
\begin{IEEEeqnarray}{rCl}
\bar	P_X& \in & \Gamma_{b}, \label{eq:b}\\
P_X & \in & \Gamma_c, \label{eq:c}\\
		P_{Y_1}& \in & \Delta_{c,c_2}.
	\end{IEEEeqnarray} 

We continue to notice  that  the upper bounds in \eqref{eq-bound-cooperative} become looser when elements are removed from the sets $\Gamma_{b}$, $\Gamma_{c}$, and $\Delta_{c,c_2}$. %or from the sets $\Gamma_{0,0}\ldots, \Gamma_{0,\W_2-1}$ and $\Gamma_{1,0}\ldots, \Gamma_{1,\W_2-1}$. 
The converse statement thus remains valid by imposing
\begin{equation}
\Delta_{c,c_2} = \{ P_{Y_1}\}.
\end{equation}
Moreover, if $b=c$, then we impose 
\begin{equation}
\Gamma_{b} =\Gamma_{c}= \{ P_X,\bar{P}_X \},
\end{equation}
and if $b \neq c$, then we impose that $\Gamma_b$ and $\Gamma_{c}$ form a partition.

If $b=c$, this concludes the proof. Otherwise, if $b \neq c$, we obtain the intermediate result that 
\begin{subequations} \label{eq:boundu cooperative larger} 
\begin{IEEEeqnarray}{rCl}
	 \theta_{1} &\leq & \min_{ \substack{\tilde P_{XY_1}: \\ \tilde P_{X} \in \Gamma_{b} ,\\ \tilde P_{Y_1}= P_{Y_1}}}  D\Big( \tilde P_{XY_1} \| \bar P_{XY_1} \Big)\\
 \theta_{2} &\leq & \min_{ \substack{\tilde P_{XY_1Y_2}: \\ \tilde P_{X} \in \Gamma_{c} ,\\ \tilde P_{Y_1}= P_{Y_1},\;  \tilde P_{Y_2}=P_{Y_2}}}  D\Big( \tilde P_{XY_1Y_2} \| \bar P_{XY_1Y_2} \Big)
\end{IEEEeqnarray}
\end{subequations}
for two sets $\Gamma_b$ and $\Gamma_c$ forming a partition of
 $\mathcal{P}(\mathcal{X})$   and satisfying \eqref{eq:b} and \eqref{eq:c}. 
 
We now characterize  the choice of the sets $\{\Gamma_b,\Gamma_c\}$ that yields the loosest bound in  \eqref{eq:boundu cooperative larger}. To this end, notice first that by assumption \eqref{eq:r}, constraints \eqref{eq:boundu cooperative larger} are equivalent to:
\begin{IEEEeqnarray}{rCl} \label{constrained-on-1}
	\theta_1 &\leq&  \min \Bigg \{  \min_{ \substack{\tilde P_{XY_1}: \\ \tilde P_{X} \in \Gamma_{b},\\  \tilde P_{Y_1}=\bar P_{Y_1}}}  D\Big( \tilde P_{XY_1} \| \bar P_{XY_1} \Big) ,\;  \min_{ \substack{\tilde P_{XY_1Y_2}: \\ \tilde P_{X} \in \Gamma_{c} ,\\ \tilde P_{Y_1}= P_{Y_1},  \tilde P_{Y_2}=P_{Y_2}}}  D\Big( \tilde P_{XY_1Y_2} \| \bar P_{XY_1Y_2} \Big) -r  \Bigg  \}.
\end{IEEEeqnarray}
We notice that the right-hand side of \eqref{constrained-on-1} is upper bounded as:
%First consider the case $b(0) \neq b(1)$. For every $\pi_X  \in \mc P(\mc X) \backslash \{P_X , \bar  P_X \}$, if $\pi_X \in \phi_{b(1)}$:
%\begin{equation} \label{constrained-on-1-p-in-j-step-1}
%	\theta_1 \leq  \min_{ \substack{\tilde P_{XY_1}: \\ \tilde P_{X} =\pi_X ,\\  \tilde P_{Y_1}=\bar P_{Y_1}}}  D\Big( \tilde P_{XY_1} \| P_{XY_1} \Big) \nonumber
%\end{equation}
%and if $\pi_X \in \phi_{b(0)}$ and then:
%\begin{equation}
% \theta_1 \leq  \min_{ \substack{\tilde P_{XY_1Y_2}: \\ \tilde P_{X} = \pi_X ,\\ \tilde P_{Y_1}= P_{Y_1},  \tilde P_{Y_2}=P_{Y_2}}}  D\Big( \tilde P_{XY_1Y_2} \| \bar P_{XY_1Y_2} \Big) -r .\nonumber
%\end{equation}
%So, for any $\pi_x \in \mc P(X) \backslash \{P_X,\bar P_X\}$
%\begin{equation}
%\theta_1 \leq \max \Bigg \{ \min_{ \substack{\tilde P_{XY_1}: \\ \tilde P_{X} =\pi_X ,\\  \tilde P_{Y_1}= \bar P_{Y_1}}}  D\Big( \tilde P_{XY_1} \| \bar P_{XY_1} \Big) ; \min_{ \substack{\tilde P_{XY_1Y_2}: \\ \tilde P_{X} = \pi_X ,\\ \tilde P_{Y_1}= P_{Y_1},  \tilde P_{Y_2}=P_{Y_2}}}  D\Big( \tilde P_{XY_1Y_2} \| P^{(2)}_{XY_1Y_2} \Big) -r \Bigg  \},
%\end{equation}
%and thus:
\begin{IEEEeqnarray}{rCl} \label{constrained-on-1-p-in-j-step-3}
\lefteqn{\min \Bigg \{  \min_{ \substack{\tilde P_{XY_1}: \\ \tilde P_{X} \in \Gamma_{b},\\  \tilde P_{Y_1}=\bar P_{Y_1}}}  D\Big( \tilde P_{XY_1} \| \bar P_{XY_1} \Big) ,\;  \min_{ \substack{\tilde P_{XY_1Y_2}: \\ \tilde P_{X} \in \Gamma_{c} ,\\ \tilde P_{Y_1}= P_{Y_1},  \tilde P_{Y_2}=P_{Y_2}}}  D\Big( \tilde P_{XY_1Y_2} \| \bar P_{XY_1Y_2} \Big) -r  \Bigg  \} } \quad \\ 
&\leq&  \min_{\pi_X  \in \mc P(\mc X) \backslash \{P_X ,\bar P_X \} } \max \Bigg \{ \min_{ \substack{\tilde P_{XY_1}: \\ \tilde P_{X} =\pi_X ,\\  \tilde P_{Y_1}= \bar P_{Y_1}}}  D\Big( \tilde P_{XY_1} \| \bar P_{XY_1} \Big) ; \min_{ \substack{\tilde P_{XY_1Y_2}: \\ \tilde P_{X} = \pi_X ,\\ \tilde P_{Y_1}= P_{Y_1},  \tilde P_{Y_2}=P_{Y_2}}}  D\Big( \tilde P_{XY_1Y_2} \| P_{XY_1Y_2} \Big) -r \Bigg  \}, 
\end{IEEEeqnarray}
and that the bound holds with equality when
\begin{equation} 
\left(\pi_X \in \Gamma_b  \right) \Longleftrightarrow \quad \left(\min_{ \substack{\tilde P_{XY_1}: \\ \tilde P_{X} =\pi_X ,\\  \tilde P_{Y_1}= \bar P_{Y_1}}}  D\Big( \tilde P_{XY_1} \| \bar P_{XY_1} \Big) \geq  \min_{ \substack{\tilde P_{XY_1Y_2}: \\ \tilde P_{X} = \pi_X ,\\ \tilde P_{Y_1}= P_{Y_1},  \tilde P_{Y_2}=P_{Y_2}}}  D\Big( \tilde P_{XY_1Y_2} \| P_{XY_1Y_2} \Big) -r \right).
\end{equation}
%we notice that \eqref{constrained-on-1-p-in-j-step-3} can be rewritten as
%\begin{IEEEeqnarray}{rCl}
%	\theta_1 &\leq&  \min_{\substack{\tilde P_{XY_1}: \\[.2ex] \tilde P_{X} \in \Gamma_{b(1)} \\[.2ex] %=\tilde{P}_X\\  
%		\tilde P_{Y_1}=P_{Y_1}^{(m)} }}% \\ (\tilde{P}_X,m) \in \bigcup_{i=1}^\mathsf{W}{\phi_{i,X} \times \mc B_i \backslash \{k\} } \end{array}}
%\!\!\!{D\Big(\tilde P_{XY_1}\|P_{XY_1}\Big)}, \nonumber \\
%\theta_2 &\leq&    \min_{\substack{\tilde P_{XY_1Y_2}\colon \\ \tilde P_{X} \in \Gamma_{b(0)},\\  \tilde P_{Y_1}\!=\!P_{Y_1}^{(m)} ,  
%		\tilde P_{Y_2}\!\!=\!P_{Y_2}^{(m)}  }}
%\!\! \hspace{-.5cm} {D\Big(\tilde P_{XY_1Y_2}\| \bar P_{XY_1Y_2}\Big)}.\nonumber
%\end{IEEEeqnarray}
%Where $(\Phi_0^*(r),\Phi_1^*(r))$ are define given \eqref{rule-partition-creation}. Now consider the case $b(0)=b(1)$. Following similar steps as above, it can be shown that it is optimal to assign all $\pi \in \mc P(X) \backslash \{P_X,\bar P_X\}$ to the same $\Phi^*_{1-b(0)}(r)$. 
This concludes the proof also for the case $b \neq c$.

\section{Proof of Converse Part of Theorem~\ref{theorem-lower-bounds-power-exponents-test-against-independence-positive-rates}} \label{App-E}

Let $R_2=0$.
Fix a rate $R_1\geq 0$  and a pair of exponents $(\theta_1,\theta_2)\in\mathcal{E}_0(R_1,0)$. Then, choose an $\epsilon \in(0,1/2)$, a sufficiently large blocklength $n$,   %\epsilon_1, \epsilon_2\in(0,1)$. Let
 encoding and decision functions $\phi_{1,n}$, $\phi_{2,n}$,   $\psi_{1,n}$, and $\psi_{2,n}$ that satisfy%~\eqref{eq:phi1}, \eqref{eq:phi2},~\eqref{eq:psi1}, and~\eqref{eq:psi2}. Let $\alpha_{1,n}$, $\alpha_{2,n}$, $\beta_{1,n}$, and $\beta_{2,n}$ be the error probabilities corresponding to the chosen functions.
  \begin{IEEEeqnarray}{rCl}
  	\alpha_{1,n} &\leq& \epsilon ,\\
  	\alpha_{2,n}& \leq &\epsilon,
  \end{IEEEeqnarray} 
  and
  \begin{IEEEeqnarray}{rCl}
  	-\frac{1}{n} \log \beta_{1,n} \geq \theta_1-\epsilon, \\
  	-\frac{1}{n} \log \beta_{2,n} \geq \theta_2-\epsilon.
  \end{IEEEeqnarray}

Notice first that for each $i\in\{1,2\}$ \cite{XK12}:
\begin{align}
D\big( P_{  \hat{\mc H}_i | H } || P_{ \hat{\mc H}_i | {\mc H} =1}\big) &= - h_2\left(\alpha_{i,n} \right) - \left(1 - \alpha_{i,n} \right) \log{\left(\beta_{i,n} \right)} \nonumber \\
& \quad - \alpha_{i,n} \log{\left(1-\beta_{i,n} \right) }
\label{proof-converse-typeII-error-independent-side-information-step1}
\end{align}
where $h_2\left(p\right)$ denotes the entropy of a Bernouilli-$(p)$ memoryless source. Since $\alpha_{i,n} \leq \epsilon<1/2$, for each $i\in\{1,2\}$,  Inequality~\eqref{proof-converse-typeII-error-independent-side-information-step1}  yields:
\begin{IEEEeqnarray}{rl}
 - \frac{1}{n}\log{\left(\beta_{i,n} \right)} 
	\leq \frac{1}{n(1-\epsilon)} D\big( P_{  \mathcal{\hat{H}}_i | \mathcal{H} = 0} || P_{ \mathcal{\hat{H}}_i | {\mc H}  = 1 }\big) + \mu_{n} \nonumber
\end{IEEEeqnarray}
with  $\mu_{n}\triangleq\frac{1}{n(1-\epsilon)} h_2\left(\epsilon\right)$. Notice that $\mu_{n}\rightarrow 0$ as $n \rightarrow \infty$.  \\[0.1ex]

Consider now:
\begin{align}
\label{eq: Decision I II error}
\theta_1 - \epsilon &\leq  - \frac{1}{n}\log{\left(\beta_{1,n} \right)}  \\&\leq  \frac{1}{n(1-\epsilon)}  D\big( P_{  \hat{\mc H}_1 | \mathcal{H}=0} || P_{ \hat{\mc H}_ | {\mc H}=1 }\big) + \mu_{n}  \\
&\stackrel{(a)}{\leq}  \frac{1}{n(1-\epsilon)} D\big( P_{Y^n_1 M_1 | {\mc H}=0}|| P_{{Y}^n_1{M_1}|{{\mc H}=1}} \big)  + \mu_{n} \\
&\stackrel{(b)}{=}  \frac{1}{n(1-\epsilon)}  I\left(  Y^n_1;M_1  \right)  + \mu_{n} \\
&\stackrel{(c)}{=}  \frac{1}{n(1-\epsilon)}\sum_{k=1}^n{H\big(  {Y_1}_k \big) - H\big(  {Y_1}_k  | M_1  {Y_1}^{k-1} \big)}   + \mu_{n} \\
&\stackrel{(d)}{\leq} \frac{1}{n(1-\epsilon)} \sum_{k=1}^n{H\big(  {Y_1}_k \big) - H\big(  {Y_1}_k  | M_1 {Y_1}^{k-1} {X}^{k-1} \big)}  + \mu_{n} \\
&\stackrel{(e)}{=}  \frac{1}{n(1-\epsilon)}  \sum_{k=1}^n{H\big(  {Y_1}_k \big) - H\big(  {Y_1}_k  |M_1 {X}^{k-1} \big)}  + \mu_{n}  \\
&\stackrel{(f)}{=}   \frac{1}{n(1-\epsilon)} \sum_{k=1}^n{I\left(  {Y_1}_k  ; U_k \right) }  + \mu_{n} \\
&\stackrel{(g)}{=} \frac{1}{n(1-\epsilon)} {I\left(  {Y_1}_Q  ; U_Q\big| Q\right) }  + \mu_{n} \\
&\stackrel{(h)}{=}  \frac{1}{1-\epsilon} {I\left(  Y_1(n) ; U(n)    \right) }  + \mu_{n}  
\end{align}
where: $(a)$ follows by the data processing inequality for  relative entropy; $(b)$ holds since $M_1$ and $Y^n_1$ are independent under the alternative hypothesis $\mc H=1$. $(c)$ is due to the chain rule for mutual information; $(d)$ follows since conditioning reduces entropy; $(e)$ is due to the Markov chain ${Y_1}^{k-1} \mkv (M_1, X^{k-1}) \mkv {Y_1}_k $; $(f)$ holds by defining $U_k \triangleq(M_1, {X}^{k-1})$; $(g)$ is obtained by introducing a  random variable $Q$ that is uniform over the set $\left\{ 1,\cdots, n \right\}$ and independent of all previously defined random variables; and $(h)$ holds by defining $U(n) \triangleq(U_Q, Q)$ and $Y_{1}(n) \triangleq Y_{1Q}$.

In a similar way, one obtains:
\begin{align}
\theta_2 - \epsilon& \leq   - \frac{1}{n}\log{\left(\beta_{2,n} \right)}\\ &\stackrel{(i)}{\leq} \frac{1}{n(1-\epsilon)}  D\big(  P_{Y^n_2 M_1 M_2 |{\mc H  =0 }}||P_{ {Y}^n_2{M}_1{M}_2|\mc H=1}\big)  + \mu_{n}  \\
&\stackrel{(j)}{=} \frac{1}{n(1-\epsilon)}  \big(I\left(  Y^n_2 ; M_1, M_2 \right)  +D(  P_{ M_1 M_2 |{\mc H=0}}||P_{{M}_1{M}_2|{ \mc H=1}})  \big ) + \mu_{n} \\
&\stackrel{(k)}{\leq} \frac{1}{n(1-\epsilon)} \big(I\left(  Y^n_2 ; M_1 \right) + I\left(  Y^n_2 ; M_2 | M_1 \right) +  D(  P_{Y^n_1 M_1  |\mc H=0}||P_{ {Y}^n_1{M}_1| \mc H=1}) \big)   + \mu_{n}  \\
&\stackrel{(\ell)}{\leq}\frac{1}{n(1-\epsilon)} \big( I(  Y^n_2; M_1)+  \log \W_2 +D(  P_{Y^n_1 M_1  |{\mc H=0}}||P_{ Y^n_1{M}_1|\mc H=1})\big)  + \mu_{n} \\
&\stackrel{(m)}{=} \frac{1}{n(1-\epsilon)} \left( I\left(  Y^n_2; M_1 \right) + I\left(  {Y^n}_1; M_1 \right)  \right)+ \tilde{\mu}_{n}   \\
&\stackrel{(o)}{\leq}\frac{1}{1-\epsilon}  \left( I\left(  Y_{2}(n); U(n) \right) +I \left(  Y_{1}(n); U(n) \right) \right)+ \tilde{\mu}_{n},
\end{align}
where $(i)$ follows by the data processing inequality for relative entropy; $(j)$ holds by the independence of the pair $(M,M_2)$ with ${Y^n}_2$ under the alternative hypothesis $\mc H=1$; $(k)$ by the data processing inequality for relative entropy; $(\ell)$ holds since conditioning reduces entropy; $(o)$ follows by proceeding along the steps $(b)$ to $(h)$ above; and $(m)$ holds by defining $\tilde{\mu}_{n}\triangleq {\W_{2,n}}/(n(1-\epsilon))+ \mu_{n}$. 

Notice that  by the assumption $R_1=0$, the term $1/n\log\|\W_{2,n}\| \to 0$ as $n \to \infty$. Thus, also $\tilde{\mu}_{n} \to 0$ as $n \to \infty$. 

We next lower bound the rate $R$:
\begin{align}
n R &\geq H\left(M_1\right)  \nonumber \\
&= H\left(M_1\right) - H\left(M|X^n\right) \nonumber \\
&= I\left(M_1;X^n\right) \nonumber \\
&= \sum_{k=1}^{n}{I\left(M_1;X_k|X^{k-1}\right)} \nonumber \\
&= \sum_{k=1}^{n}{I\left(X_k;U_k\right)} \nonumber \\
&= n I\left(X_Q;U_Q|Q\right) \nonumber \\
&=  n I\left(U(n) ; X(n)\right)  \nonumber
\end{align}

For any blocklength $n$, the newly defined random variables $X(n),Y_{1}(n),Y_{2}(n) \sim P_{X Y_1 Y_2}$ and $U(n) \mkv X(n) \mkv (Y_{1}(n),Y_{2}(n))$.
 Letting now the blocklength $n\to \infty$, and then $\epsilon \to 0$, by continuity of mutual information establishes the desired converse result. %   the asymptotic exponents 
%\begin{IEEEeqnarray}{rCl}
%	\theta_{1}& \triangleq&  \varliminf_{n \to \infty} \theta_{1,n}\\
%	\theta_{2}& \triangleq&  \varliminf_{n \to \infty} \theta_{2,n}  
%\end{IEEEeqnarray}
%satisfy
%\begin{align}
%\theta_1 &\leq I\left(U ; Y_1\right)\\
%\theta_2 &\leq I\left(U ; Y_1\right) +  I\left(U ; Y_2\right);
%\end{align}
%for some  $U \in \mathcal{S}_{\text{nc}}\left(R_1\right)$.
%This completes the proof of converse part of proposition~\ref{theorem-lower-bounds-power-exponents-test-against-independence-positive-rates}.

\section{Proof of Theorem~\ref{theorem-lower-bounds-power-exponents-general-hypotheses-positive-rates}}~\label{secV_subsecA}
We analyze the scheme in Subsection~\ref{sec:coding}, where we  focus on the type-II error probabilities. The analysis of the type-I error  probabilities is standard and  omitted.

To analyze the probability of type-II error at  Detector 1, we notice that $\hat{\mathcal{H}}_1=0$ only if there exists an index $ m_1 \in \{1,\ldots, 2^{nR_1}\}$ such that 
\begin{equation}
 (U^n(m_1),X^n)\in \mathcal{T}_{\mu/8}^n(P_{UX})\quad \textnormal{and} \quad  (U^n(m_1),Y_1^n)\in \mathcal{T}_{\mu/4}^n(P_{UY_1}) 
\end{equation} 
Therefore, using the union bound: 
	\begin{align}
	&\Pr \left[\hat{\mathcal{H}}_1=0| \mathcal{H}=1\right] \nonumber\\
%%%
%	&\leq \Pr \left[\exists m \in \{1,\ldots, 2^{nR}\} \colon (U^n(m),X^n)\in \mathcal{T}_{\mu/8}^n(P_{UX}),\;\; (U^n(m),Y^n)\in \mathcal{T}_{\mu}^n(P_{UY} \; | \mathcal{H}=1\right] \nonumber\\	
	%%%
	&\leq\sum_{m_1=1}^{2^{nR_1}} \Pr \left[(U^n(m_1),X^n)\in \mathcal{T}_{\mu/8}^n(P_{UX}),\;\; (U^n(m),Y_1^n)\in \mathcal{T}_{\mu/4}^n(P_{UY_1})\; \Big|\; \mathcal{H}=1 \right]\\
	%%%
%	&\leq\sum_{m} \Pr \left[(U^n(m),X^n)\in \mathcal{T}_{\mu/2}^n(P_{UX}),\;\; (U^n(m),Y^n)\in \mathcal{T}_{\mu}^n(P_{UY})\;  \Big|\; \mathcal{H}=1  \right]\nonumber\\
	%%%
	&\stackrel{(a)}{\leq} 2^{nR_1} \cdot \max_{\substack{\tilde{P}:\\ |\tilde{P}_{UX}-P_{UX}|<\mu/8\\ |\tilde{P}_{UY_1}-P_{UY_1}|<\mu/4}} 2^{-n (D(\tilde{P}_{UXY_1}\|P_U\bar{P}_{XY_1} )- \xi_n)}\label{eq:prob2}\\
	& \stackrel{(b)}{<} \max_{\substack{\tilde{P}:\\ |\tilde{P}_{UX}-P_{UX}|<\mu/8\\ |\tilde{P}_{UY_1}-P_{UY_1}|<\mu/4}} 2^{n( I(U;X)-  D(\tilde{P}_{UXY_1}\|P_U\bar{P}_{XY_1})- \xi_n)}  \\
	& = \max_{\substack{\tilde{P}:\\ |\tilde{P}_{UX}-P_{UX}|<\mu/8\\ |\tilde{P}_{UY_1}-P_{UY_1}|<\mu/4}}
	 2^{-n( D(\tilde{P}_{UXY_1}\|P_{U|X}\bar{P}_{XY_1})- \xi_n)}  ,
	\end{align}
where $x_n$ is a sequence that tends to $0$ as $n\to \infty$.  Inequality $(a)$ holds by Sanov's theorem and by the way the source sequences and the codewords are generated and Inequality $(b)$ holds by the choice of $R_1$ in \eqref{eq:rate1}.
	
To analyze the probability of type-II error at Detector 2, we notice that $\hat{\mathcal{H}}_2=0$ only if there exists a pair of indices $(m_1,m_2)\in \{1,\ldots, 2^{nR_1}\}\times \{1,\ldots, 2^{nR_2}\}$ so that 
\begin{IEEEeqnarray}{rCl}
&(U^n(m_1),X^n)\in \mathcal{T}_{\mu/8}^n(P_{UX})\quad \textnormal{and} \quad  &(U^n(m_1), V^n(m_2|m_1),Y_1^n)\in \mathcal{T}_{\mu/2}^n(P_{UVY_1})\nonumber \\
& &\hspace{2cm} \textnormal{and} \quad  (U^n(m_1),V^n(m_2|m_1),Y_2^n)\in \mathcal{T}_{\mu}^n(P_{UVY_2}) 
\end{IEEEeqnarray} 
Therefore, applying similar steps as before:
	\begin{align}
	&\Pr \left[\hat{\mathcal{H}}_2=0| \mathcal{H}=1\right] \nonumber\\
	%%%
	&\leq \sum_{m_1=1}^{2^{nR_1}}  \sum_{m_2=1}^{2^{nR_2}}  \Pr \Big[(U^n(m_1),X^n)\in \mathcal{T}_{\mu/8}^n(P_{UX}),\;  (U^n(m_1), V^n(m_2|m_1),Y_1^n)\in \mathcal{T}_{\mu/2}^n(P_{UVY_1}) , \nonumber \\
	& \hspace{6.2cm}\;  (U^n(m_1),V^n(m_2|m_1),Y_2^n)\in \mathcal{T}_{\mu/2}^n(P_{UVY_2})  \;  \Big |\; \mathcal{H}=1 \Big]\\
	&<  2^{n (I(U;X)+I(V;Y_1|U))} \cdot \max_{\substack{\tilde{P}:\\ |\tilde{P}_{UX}-P_{UX}|<\mu/8\\ |\tilde{P}_{UVY_1}-P_{UVY_1}|<\mu/2\\ |\tilde{P}_{UVY_2}-P_{UVY_2}|<\mu }}	 2^{-n( D(\tilde{P}_{UVXY_1Y_2}\|P_{U}P_{V|U} \bar{P}_{XY_2})- \xi_n')}\\
	& = \max_{\substack{\tilde{P}:\\ |\tilde{P}_{UX}-P_{UX}|<\mu/8\\ |\tilde{P}_{UVY_1}-P_{UVY_1}|<\mu/2\\ |\tilde{P}_{UVY_2}-P_{UVY_2}|<\mu/4 }}2^{-n ( D(\tilde{P}_{UVXY_1Y_2}\|P_{U|X}P_{V|UY_1} \bar{P}_{XY_1Y_2})- \xi_n')}
	\end{align}
where $\xi_n'$ is a sequence that tends to $0$ as $n\to \infty$.	
The proof is then concluded by letting $n \to \infty$ and by noting that there must exist at least one pair of codebooks achieving the same exponents as the random ensemble.

\section{Proof of Theorem~\ref{thm:concurrent}}~\label{app:concurrent}
We analyze the scheme in Subsection~\ref{sec:coding}. The type-II error probability at Detector 2 can be analyzed as in the preceding Appendix~\ref{secV_subsecA}. 

To analyze the probability of type-II error at  Detector 1, we notice that $\hat{\mathcal{H}}_1=0$ only if there exists an index $ m_1 \in \{1,\ldots, 2^{nR_1}\}$ such that 
\begin{equation}
(U_1^n(m_1),X^n)\in \mathcal{T}_{\mu/8}^n(\bar{P}_{U_1X})\quad \textnormal{and} \quad  (U_1^n(m_1),Y_1^n)\in \mathcal{T}_{\mu/4}^n(\bar{P}_{U_1Y_1}) 
\end{equation} 
By now standard arguments: 
\begin{align}
\Pr \left[\hat{\mathcal{H}}_1=0| \mathcal{H}=1\right] 
&\leq\sum_{m_1=1}^{2^{nR_1}} \Pr \left[(U_1^n(m_1),X^n)\in \mathcal{T}_{\mu/8}^n(\bar{P}_{U_1X}),\;\; (U_1^n(m),Y_1^n)\in \mathcal{T}_{\mu/4}^n(\bar{P}_{U_1Y_1})\; \Big|\; \mathcal{H}=1 \right]\\
%%%
%%	&\leq\sum_{m} \Pr \left[(U^n(m),X^n)\in \mathcal{T}_{\mu/2}^n(P_{UX}),\;\; (U^n(m),Y^n)\in \mathcal{T}_{\mu}^n(P_{UY})\;  \Big|\; \mathcal{H}=1  \right]\nonumber\\
%%%%
%&\stackrel{(a)}{\leq} 2^{nR_1} \cdot \max_{\substack{\tilde{P}:\\ |\tilde{P}_{UX}-P_{UX}|<\mu/8\\ |\tilde{P}_{UY_1}-P_{UY_1}|<\mu/4}} 2^{-n (D(\tilde{P}_{UXY_1}\|P_U\bar{P}_{XY_1} )- \xi_n)}\label{eq:prob2}\\
& <\max_{\substack{\tilde{P}:\\ |\tilde{P}_{\bar{U}_1X}-\bar{P}_{{U}_1X}|<\mu/8\\ |\tilde{P}_{{U}_1Y_1}-\bar{P}_{{U}_1Y_1}|<\mu/4}}
2^{-n( D(\tilde{P}_{\bar{U}_1XY_1}\|{P}_{U_1|X}{P}_{XY_1})- \xi_n'')},
\end{align}
where the sequence $\xi_n'' \to 0$ as $n \to \infty$.

%------------------------------------------------------------------------------------------------------
%--------------------------------------------------------------------------------------------------===================
%
\bibliographystyle{IEEEtran}
\bibliography{IEEEabrv,paper2018}
\end{document}